\documentclass[usenatbib]{mnras}
\usepackage{graphicx}
\usepackage{amsmath}
\usepackage{amssymb}
\usepackage{color}
\usepackage{url}
\usepackage{CJKutf8}

\usepackage{threeparttable}

\newcommand\lsim{\mathrel{\rlap{\lower4pt\hbox{\hskip1pt$\sim$}}
        \raise1pt\hbox{$<$}}}
\newcommand\gsim{\mathrel{\rlap{\lower4pt\hbox{\hskip1pt$\sim$}}
        \raise1pt\hbox{$>$}}}
\newcommand{\lya}{\ifmmode\mathrm{Ly}\alpha\else{}Ly$\alpha$\fi}
\newcommand{\lyb}{\ifmmode\mathrm{Ly}\beta\else{}Ly$\beta$\fi}
\newcommand{\igm}{\ifmmode\mathrm{IGM}\else{}IGM\fi}
\newcommand{\lae}{\ifmmode\mathrm{LAE}\else{}LAE\fi}
\newcommand{\h}{\ifmmode\mathrm{H}\else{}H\fi}
\newcommand{\hi}{\ifmmode\mathrm{H\,{\scriptscriptstyle I}}\else{}H\,{\scriptsize I}\fi}
\newcommand{\hii}{\ifmmode\mathrm{H\,{\scriptscriptstyle II}}\else{}H\,{\scriptsize II}\fi}
\newcommand{\cmb}{\ifmmode\mathrm{CMB}\else{}CMB\fi}
\newcommand{\qso}{\ifmmode\mathrm{QSO}\else{}QSO\fi}
\newcommand{\eor}{\ifmmode\mathrm{EoR}\else{}EoR\fi}
\newcommand{\heii}{\ifmmode\mathrm{He\,{\scriptscriptstyle II}}\else{}He\,{\scriptsize II}\fi}
\newcommand{\heiii}{\ifmmode\mathrm{He\,{\scriptscriptstyle III}}\else{}He\,{\scriptsize III}\fi}
\newcommand{\ciii}{\ifmmode\mathrm{C\,{\scriptscriptstyle III]}}\else{}C\,{\scriptsize III]}\fi}
\newcommand{\oiii}{\ifmmode\mathrm{O\,{\scriptscriptstyle III}}\else{}O\,{\scriptsize III}\fi}
\newcommand{\aliii}{\ifmmode\mathrm{Al\,{\scriptscriptstyle III}}\else{}Al\,{\scriptsize III}\fi}
\newcommand{\mgii}{\ifmmode\mathrm{Mg\,{\scriptscriptstyle II}}\else{}Mg\,{\scriptsize II}\fi}
\newcommand{\fe}{\ifmmode\mathrm{Fe}\else{}Fe\fi}
\newcommand{\nv}{\ifmmode\mathrm{N\,{\scriptscriptstyle V}}\else{}N\,{\scriptsize V}\fi}
\newcommand{\niv}{\ifmmode\mathrm{N\,{\scriptscriptstyle IV]}}\else{}N\,{\scriptsize IV]}\fi}
\newcommand{\cii}{\ifmmode\mathrm{C\,{\scriptscriptstyle II}}\else{}C\,{\scriptsize II}\fi}
\newcommand{\civ}{\ifmmode\mathrm{C\,{\scriptscriptstyle IV}}\else{}C\,{\scriptsize IV}\fi}
\newcommand{\siv}{\ifmmode\mathrm{Si\,{\scriptscriptstyle IV}}\else{}Si\,{\scriptsize IV}\fi}
\newcommand{\siii}{\ifmmode\mathrm{Si\,{\scriptscriptstyle II}}\else{}Si\,{\scriptsize II}\fi}
\newcommand{\siiii}{\ifmmode\mathrm{Si\,{\scriptscriptstyle III]}}\else{}Si\,{\scriptsize III]}\fi}
\newcommand{\ovi}{\ifmmode\mathrm{O\,{\scriptscriptstyle VI}}\else{}O\,{\scriptsize VI}\fi}
\newcommand{\sioiv}{\ifmmode\mathrm{Si\,{\scriptscriptstyle IV}\,\plus O\,{\scriptscriptstyle IV]}}\else{}Si\,{\scriptsize IV}\,+O\,{\scriptsize IV]}\fi}

\newcommand{\avenf}{$\bar{x}_{\rm HI}$}

\newcommand{\cmfst}{\textsc{\small 21CMFAST}}

\pdfoutput=1
\title[WST 21-cm non-Gaussianity]{Detecting the non-Gaussianity of the 21-cm signal during reionisation with the Wavelet Scattering Transform}
\author[B. Greig et al.] {Bradley~Greig$^{1,2}$\thanks{E-mail:~greigb@unimelb.edu.au}, Yuan-Sen Ting (丁源森)$^{2,3,4}$ \& Alexander A. Kaurov$^{5, 6, 7}$ \\
$^1$School of Physics, University of Melbourne, Parkville, VIC 3010, Australia \\
$^2$ARC Centre of Excellence for All-Sky Astrophysics in 3 Dimensions (ASTRO 3D) \\
$^3$Research School of Astronomy \& Astrophysics, Australian National University, Cotter Rd., Weston, ACT 2611, Australia \\
$^4$School of Computing, Australian National University, Acton ACT 2601, Australia \\
$^5$Department of the History of Science, Harvard University, Cambridge, MA, USA \\
$^6$Program in Interdisciplinary Studies, Institute for Advanced Study, Princeton, NJ, USA \\
$^7$Blue Marble Space Institute of Science, Seattle, WA, USA \\
}

\begin{document}
\label{firstpage}
\pagerange{\pageref{firstpage}--\pageref{lastpage}}
\begin{CJK}{UTF8}{gkai} 
\maketitle
\end{CJK}

\begin{abstract}
\noindent
Detecting the 21-cm hyperfine transition from neutral hydrogen in the intergalactic medium is our best probe for understanding the astrophysical processes driving the Epoch of Reionisation (EoR). The primary means for a detection of this 21-cm signal is through a statistical measurement of the spatial fluctuations using the 21-cm power spectrum (PS). However, the 21-cm signal is non-Gaussian meaning the PS, which only measures the Gaussian fluctuations, is sub-optimal for characterising all of the available information. The upcoming Square Kilometre Array (SKA) will perform a deep, 1000 hr observation over 100 deg$.^{2}$ specifically designed to recover direct images of the 21-cm signal. In this work, we use the Wavelet Scattering Transform (WST) to extract the non-Gaussian information directly from these two-dimensional images of the 21-cm signal. The key advantage of the WST is its stability with respect to statistical noise for measuring non-Gaussian information, unlike the bispectrum whose statistical noise diverges. We introduce a novel method to isolate this non-Gaussian information from mock 21-cm images and demonstrate its detection at 150 (177)~MHz ($z\sim8.5$ and $\sim7$) for a fiducial model with signal-to-noise of $\sim$5~(8) assuming perfect foreground removal and $\sim2$~(3) assuming foreground wedge avoidance.
\end{abstract} 

\section{Introduction}

The early Universe, following recombination, is invisible to most forms of radiation due to the omnipresent neutral hydrogen fog. Gradually, this fog is lifted from the intergalactic medium (IGM) via the ionisation of the neutral hydrogen by escaping ultra-violet (UV) photons from the first stars and galaxies. This process, the final baryonic phase transition of the Universe, is referred to as the Epoch of Reionisation (EoR).

Our most promising approach to study the EoR is to observe the spatial distribution of neutral hydrogen in the IGM and how it disappears over cosmic time. This can be achieved by detecting the intensity of 21-cm photons emitted by the neutral hydrogen through the hyperfine spin-flip transition in contrast against a background radiation source, such as the cosmic microwave background \citep[CMB; see e.g.][]{Gnedin:1997p4494,Madau:1997p4479,Shaver:1999p4549,Tozzi:2000p4510,Gnedin:2004p4481,Furlanetto:2006p209,Morales:2010p1274,Pritchard:2012p2958}. This 21-cm signal can be observed either in emission or absorption, depending on the thermal and ionisation state of the IGM. Since it is a line transition, observing it at different frequencies (redshifts) builds a three-dimensional movie outlining the evolution of the IGM during the early Universe. This detailed picture highlighting the spatial morphology of the IGM enables us to infer the properties of the stars and galaxies responsible for the EoR.

To observe the spatial fluctuations in the 21-cm signal we require large-scale interferometer experiments. The first generation of these, with their lower sensitivities and smaller collecting areas such as the Low-Frequency Array (LOFAR; \citealt{vanHaarlem:2013p200}), the Murchison Wide Field Array (MWA; \citealt{Tingay:2013p2997,Wayth:2018}), the Precision Array for Probing the Epoch of Reionisation (PAPER; \citealt{Parsons:2010p3000}), the Owens Valley Radio Observatory Long Wavelength Array (OVRO-LWA; \citealt{Eastwood:2019}) and the upgraded Giant Metrewave Radio Telescope (uGMRT; \citealt{Gupta:2017}) are primarily designed to yield a low signal-to-noise detection of the statistical properties of the spatial fluctuations through the 21-cm power spectrum (PS). However, the 21-cm signal, due to its complex three dimensional morphology is non-Gaussian meaning the PS is a sub-optimal statistic for understanding the EoR. Only with next generation experiments, such as the Hydrogen Epoch of Reionization Array (HERA; \citealt{DeBoer:2017p6740}), NenuFAR (New extension in Nan\c{c}ay Upgrading loFAR; \citealt{Zarka:2012}) and the Square Kilometre Array (SKA; \citealt{Mellema:2013p2975,Koopmans:2015}) will we have the capacity to yield significantly higher sensitivities enabling the non-Gaussian properties of the 21-cm signal to be explored.

While non-Gaussianity can be explored with higher order moments such as the 21-cm bispectrum \citep[e.g.][]{Yoshiura:2015,Shimabukuro:2016,Majumdar:2018,Watkinson:2019,Hutter:2020,Majumdar:2020,Kamran:2021}, these suffer from rapidly increasing statistical variance and are sensitive to outliers in the data. Importantly, the SKA is also designed to yield tomographic images of the 21-cm signal \citep{Mellema:2013p2975} which adds an entirely new dimension in how we can explore the EoR beyond just statistical estimators of the spatial fluctuations in Fourier space. For example, one-point statistics of 21-cm signal \citep{Watkinson:2014,Shimabukuro:2015,Kubota:2016,Banet:2021,Gorce:2021}, large-scale morphological and/or topological features \citep[e.g.][]{Yoshiura:2017,Bag:2019,Chen:2019,Elbers:2019,Kapahtia:2019,Gazagnes:2021,Giri:2021,Kapahtia:2021} or the distribution of ionised regions \citep{Kakiichi:2017,Giri:2018a,Giri:2018b,Giri:2019a,Bianco:2021}. Deep learning with convolutional neural networks (CNNs) have also been applied to the expected 3D 21-cm signal \citep[e.g.][]{Doussot:2019,Gillet:2019,Hassan:2019,LaPlante:2019,Hassan:2020,Kwon:2020,Mangena:2020,Prelogovic:2021,Zhao:2022}.

In \citet{Greig:WST} we introduced an alternative method called the Wavelet Scattering Transform (WST), capable of extracting non-Gaussian information directly from 2D images of the 21-cm signal. The WST, first introduced by \citet{Mallat:2012}, shares several properties with CNNs, primarily the convolution of the input image by a family of wavelet filters to extract spatial features on different physical scales. After each successive convolution, the modulus of the filtered image is taken to ensure stability in the statistical properties (i.e. prevents the noise from diverging). Finally, the filtered image is spatially averaged to compress the information into a single number, a scattering coefficient, for each spatial scale. The ease in applying and interpreting the WST, has already seen it used across several topics in astronomy; the interstellar medium \citep{Allys:2019,Blancard:2020,Saydjari:2021}, weak lensing \citep{Cheng:2020,Cheng:2021} and large-scale structure \citep{Allys:2020,Valogiannis:2021,Valogiannis:2022}.

Our previous work focussed on applying the WST for astrophysical parameter inference using Fisher Matrices, demonstrating that the WST outperforms the 21-cm PS due to its ability to measure the non-Gaussian information. In this work we are instead specifically interested in the non-Gaussian signal, introducing a novel method to more cleanly isolate the non-Gaussian information from 2D images of the 21-cm signal. We then demonstrate the application of this method to the planned SKA imaging survey, highlighting the strength to which the non-Gaussian signal can be detected from a fiducial astrophysical model. The ease in which we can isolate the non-Gaussian signal from our approach could provide the first ever detection of the non-Gaussianity in the 21-cm signal during reionisation.

This paper is organised as follows. In Section~\ref{sec:Method} we summarise our 21-cm simulations along with our prescription for constructing realistic 21-cm images in the presence of instrumental and astrophysical foreground effects. Next, in Section~\ref{sec:Understanding} we outline the WST and introduce our method for isolating the non-Gaussian information embedded in the 21-cm images. In Section~\ref{sec:results} we apply our method to the planned SKA imaging survey, discussing the main results. In Section~\ref{sec:discussion} we provide a brief discussion before concluding with our final remarks in Section~\ref{sec:conclusion}. Unless explicitly mentioned, all quantities are expressed in co-moving units and we adopt the cosmological parameters:  ($\Omega_\Lambda$, $\Omega_{\rm M}$, $\Omega_b$, $n$, $\sigma_8$, $H_0$) = (0.69, 0.31, 0.048, 0.97, 0.81, 68 km s$^{-1}$ Mpc$^{-1}$), consistent with recent results from the Planck mission \citep{Planck:2020}.

\section{Simulating the 21-cm signal} \label{sec:Method}

\subsection{\cmfst{}}

To simulate the cosmic 21-cm signal we make use of the computationally efficient semi-numerical code \cmfst{}\footnote{https://github.com/21cmfast/21cmFAST}\citep{Mesinger:2007p122,Mesinger:2011p1123,Murray:2020}. Specifically we generate three dimensional light-cones of the 21-cm brightness temperature signal according to,
\begin{eqnarray} \label{eq:21cmTb}
\delta T_{\rm b}(\nu) &=& \frac{T_{\rm S} - T_{\rm CMB}(z)}{1+z}\left(1 - {\rm e}^{-\tau_{\nu_{0}}}\right)~{\rm mK},
\end{eqnarray}
where $T_{\rm S}$ is the spin temperature, $T_{\rm CMB}$ is the CMB temperature and $\tau_{\nu_{0}}$ is the optical depth of the 21-cm line,
\begin{eqnarray}
\tau_{\nu_{0}} &\propto& (1+\delta_{\rm nl})(1+z)^{3/2}\frac{x_{\hi{}}}{T_{\rm S}}\left(\frac{H}{{\rm d}v_{\rm r}/{\rm d}r+H}\right),
\end{eqnarray}
where $x_{\hi{}}$ is the neutral hydrogen fraction, $\delta_{\rm nl} \equiv \rho/\bar{\rho} - 1$ is the gas overdensity, $H(z)$ is the Hubble parameter and ${\rm d}v_{\rm r}/{\rm d}r$ is the line-of-sight peculiar velocity gradient. Each quantity is evaluated at $z = \nu_{0}/\nu - 1$, where $\nu$ is the observing frequency, and we have dropped the spatial dependence for brevity.

\subsubsection{Ionisation state of the IGM}

\cmfst{} determines the ionisation state of the IGM using an excursion-set approach \citep[e.g.][]{Furlanetto:2004p123}, which compares the cumulative number of ionising photons, $n_{\rm ion}$ to the total number of neutral hydrogen atoms plus cumulative recombinations, $\bar{n}_{\rm rec}$ \citep[e.g.][]{Sobacchi:2014p1157} in spheres of decreasing radii. A simulation cell is considered ionised when,
\begin{eqnarray} \label{eq:ioncrit}
n_{\rm ion}(\boldsymbol{x}, z | R, \delta_{R}) \geq (1 + \bar{n}_{\rm rec})(1-\bar{x}_{e}).
\end{eqnarray}
Here, the $(1-\bar{x}_{e})$ factor accounts for ionisations by X-rays and the left hand side is the cumulative number of ionising photons per baryon inside a spherical region of size, $R$ and overdensity, $\delta_{R}$,
\begin{eqnarray} \label{eq:ioncrit2}
n_{\rm ion} = \bar{\rho}^{-1}_b\int^{\infty}_{0}{\rm d}M_{\rm h} \frac{{\rm d}n(M_{h}, z | R, \delta_{R})}{{\rm d}M_{\rm h}}f_{\rm duty} \dot{M}_{\ast}f_{\rm esc}N_{\gamma/b},
\end{eqnarray}
where $\bar{\rho}_b$ is the mean baryon density, $\frac{{\rm d}n}{{\rm d}M_{\rm h}}$ is the halo mass function (HMF)\footnote{Specifically, the Sheth-Tormen HMF \citep{Sheth:1999p2053}} and $N_{\gamma/b}$ is the total number of ionising photons per stellar baryon\footnote{We adopt $N_{\gamma/b}=5000$, consistent with a Salpeter initial mass function \citep{Salpeter:1955}.}.

The quantities $\dot{M}_{\ast}$, $f_{\rm esc}$ and $f_{\rm duty}$ parameterise the UV properties of the galaxies responsible for reionisation, specifically the star-formation rate (SFR), the escape fraction of UV photons and the duty-cycle \citep[see][for further details]{Park:2019}. In short, this model assumes the galaxy stellar mass, $M_{\ast}$, can be directly related to its host halo mass, $M_{\rm h}$,
\begin{eqnarray}
M_{\ast}(M_{\rm h}) = f_{\ast}\left(\frac{\Omega_{\rm b}}{\Omega_{\rm m}}\right)M_{\rm h},
\end{eqnarray}
with the fraction of galactic gas in stars, $f_{\ast}$, represented by a power-law in halo mass,
\begin{eqnarray}
f_{\ast} = f_{\ast, 10}\left(\frac{M_{\rm h}}{10^{10}\,M_{\odot}}\right)^{\alpha_{\ast}},
\end{eqnarray}
where $\alpha_{\ast}$ is the power-law index and $f_{\ast, 10}$ corresponds to the normalisation evaluated at a dark matter halo of mass $10^{10}$~$M_{\odot}$. Next, dividing the stellar mass by a characteristic time-scale yields an estimate of the SFR,
\begin{eqnarray} \label{}
\dot{M}_{\ast}(M_{\rm h},z) = \frac{M_{\ast}}{t_{\ast}H^{-1}(z)},
\end{eqnarray}
where $H^{-1}(z)$ is the Hubble time and $t_{\ast}\in(0.05,1]$. Like above, we also parameterise $f_{\rm esc}$ as a power-law in halo mass,
\begin{eqnarray} \label{}
f_{\rm esc} = f_{\rm esc, 10}\left(\frac{M_{\rm h}}{10^{10}\,M_{\odot}}\right)^{\alpha_{\rm esc}}.
\end{eqnarray}

Finally, the duty cycle, $f_{\rm duty}$ accounts for the fact that only a fraction of low mass haloes below some mass scale $M_{\rm turn}$ can host star-forming galaxies due to feedback and/or inefficient cooling,
\begin{eqnarray} \label{eq:duty}
f_{\rm duty} = {\rm exp}\left(-\frac{M_{\rm turn}}{M_{\rm h}}\right).
\end{eqnarray}
In total, this model contains six free parameters ($f_{\ast, 10}$, $f_{\rm esc, 10}$, $\alpha_{\ast}$, $\alpha_{\rm esc}$, $t_{\ast}$ and $M_{\rm turn}$) describing the UV properties of the galaxies responsible for reionisation.

\subsubsection{Thermal state of the IGM}

\cmfst{} determines the thermal state of the neutral IGM by self-consistently calculating the heating/cooling from structure formation, Compton scattering off CMB photons and the heating by X-rays and partial ionisations. These set the IGM spin temperature, $T_{\rm S}$, which is the weighted mean of the gas, $T_{\rm K}$, CMB, $T_{\rm CMB}$, and colour, $T_{\alpha}$, temperatures,
\begin{eqnarray} \label{}
T^{-1}_{\rm S} = \frac{T^{-1}_{\rm CMB} + x_{\alpha}T^{-1}_{\alpha} + x_{\rm c}T^{-1}_{\rm K}}{1 + x_{\alpha} + x_{\rm c}},
\end{eqnarray}
where $x_{\alpha}$ is the Wouthuysen-Field coupling coefficient \citep{Wouthuysen:1952p4321,Field:1958p1} and $x_{\rm c}$ is the collisional coupling coefficient between the free electrons and CMB photons. Determined at each simulation cell, it also depends on the local gas density and the intensity of the background Lyman-$\alpha$ (Ly$\alpha$) radiation which is the summed contribution of X-ray excitations of neutral hydrogen atoms and the direct stellar emission of Lyman band photons by the first sources. For further details, see \citet{Mesinger:2011p1123}.

X-rays emitted by stellar remnants within the first galaxies can escape the host galaxy and heat the IGM. This is modelled by the angle-averaged specific X-ray intensity, $J(\boldsymbol{x}, E, z)$, (in erg s$^{-1}$ keV$^{-1}$ cm$^{-2}$ sr$^{-1}$),
\begin{equation} \label{eq:Jave}
J(\boldsymbol{x}, E, z) = \frac{(1+z)^3}{4\pi} \int_{z}^{\infty} dz' \frac{c dt}{dz'} \epsilon_{\rm X}  e^{-\tau},
\end{equation}
calculated by integrating the co-moving X-ray specific emissivity, $\epsilon_{\rm X}(\boldsymbol{x}, E_e, z')$, back along the light-cone with the attenuation of X-rays by the IGM determined by $e^{-\tau}$. This emissivity, $\epsilon_{\rm X}$, is given by,
\begin{equation} \label{eq:emissivity}
\epsilon_{\rm X}(\boldsymbol{x}, E_{\rm e}, z') = \frac{L_{\rm X}}{\rm SFR} \left[ (1+\bar{\delta}_{\rm nl}) \int^{\infty}_{0}{\rm d}M_{\rm h} \frac{{\rm d}n}{{\rm d}M_{\rm h}}f_{\rm duty} \dot{M}_{\ast}\right],
\end{equation}
where the quantity in square brackets is the SFR density along the light-cone, $\bar{\delta}_{\rm nl}$ is the mean, non-linear density in a shell centred on the simulation cell $(\boldsymbol{x}, z)$ and $L_{\rm X}/{\rm SFR}$ (erg s$^{-1}$ keV$^{-1}$ $M^{-1}_{\odot}$ yr) is the specific X-ray luminosity per unit star formation escaping the host galaxies assuming a power-law with respect to photon energy, $L_{\rm X} \propto E^{- \alpha_X}$. We then normalise this to the integrated soft-band ($<2$~keV) luminosity per SFR (in erg s$^{-1}$ $M^{-1}_{\odot}$ yr),
\begin{equation} \label{eq:normL}
  L_{{\rm X}<2\,{\rm keV}}/{\rm SFR} = \int^{2\,{\rm keV}}_{E_{0}} dE_e ~ L_{\rm X}/{\rm SFR} ~,
\end{equation}
where $E_0$ is the threshold energy below which X-ray photons are absorbed by the host galaxy. In total, this amounts to three additional parameters describing the galaxies X-ray properties, $L_{{\rm X}<2\,{\rm keV}}/{\rm SFR}$, $E_0$ and $\alpha_X$. Throughout, we adopt $\alpha_X = 1$ consistent with high-mass X-ray binary observations in the local Universe \citep{Mineo:2012p6282,Fragos:2013p6529,Pacucci:2014p4323}.

\subsection{Astrophysical models} \label{sec:setup}

\begin{table*}
\begin{tabular}{@{}lccccccccc}
\hline
Model Type  & ${\rm log_{10}}(f_{\ast,10})$ & $\alpha_{\ast}$ & ${\rm log_{10}}(f_{\rm esc,10})$ & $\alpha_{\rm esc}$ & $t_{\ast}$ & ${\rm log_{10}}(M_{\rm turn})$ & ${\rm log_{10}}\left(\frac{L_{{\rm X}<2{\rm keV}}}{\rm SFR}\right)$ & $E_0$   \\
               &  &  &  &  & & $[{\rm M_{\sun}}]$ & $[{\rm erg\,s^{-1}\,M_{\sun}^{-1}\,yr}]$ &  $[{\rm keV}]$ \\
\hline
\vspace{0.8mm}
Fiducial Model & $-1.30$ & $0.50$ & $-1.00$ & $-0.50$ & $0.5$ & $8.7$ & $40.50$ &  $0.50$\\
\hline
\vspace{0.8mm}
Cold Reionisation &  $-1.30$ & $0.50$ & $-1.00$ & $-0.50$ & $0.5$ & $8.7$ & $38.00$ &  $0.50$  \\
\hline
\vspace{0.8mm}
Large Haloes &   $-0.70$ & $0.50$ & $-1.00$ & $-0.50$ & $0.5$ & $9.9$ & $40.50$ &  $0.50$  \\
\hline
\vspace{0.8mm}
Extended Reionisation &  $-1.65$ & $0.50$ & $-1.00$ & $-0.50$ & $0.5$ & $8.0$ & $40.50$ &  $0.50$ \\
\hline
\end{tabular}
\caption{Summary of the astrophysical parameters for the four different reionisation models. See Section~\ref{sec:setup} for further details.}
\label{tab:Models}
\end{table*} 

We consider the same four reionisation models as in \citet{Greig:WST} to gain physical insights. Table~\ref{tab:Models} provides a summary of the adopted astrophysical parameters, with a brief justification for each provided below:
\begin{itemize}
\item[1.] \textit{Fiducial Model}: the default \citet{Park:2019} model, consistent with a broad range of observational constraints\footnote{For a more up-to-date model including \lya{} forest data, see \citet{Qin:2021}.} including observed UV LFs at $z=6-10$ and the electron scattering optical depth, $\tau_{\rm e}$.\\
\item[2.] \textit{Large Haloes}: reionisation driven by larger, more biased galaxies (larger $M_{\rm turn}$) producing fewer but larger ionised regions and an increase in the amplitude of the 21-cm signal.\\
\item[3.] \textit{Cold Reionisation}: the same as our fiducial model except that the IGM undergoes little to no X-ray heating resulting in extremely large amplitude brightness temperature contrasts \citep[e.g.][]{Mesinger:2014p244,Parsons:2014p781}.\\
\item[4.] \textit{Extended Reionisation}: a slower, more extended reionisation driven by low efficiency faint star-forming galaxies which result in a larger number of small ionised regions, producing a lower amplitude 21-cm signal and a considerably later reionisation ($z\sim5$).
\end{itemize}

\begin{figure*} 
	\begin{center}
		\includegraphics[trim = 3.3cm -0.2cm 0cm 0.5cm, scale = 0.288]{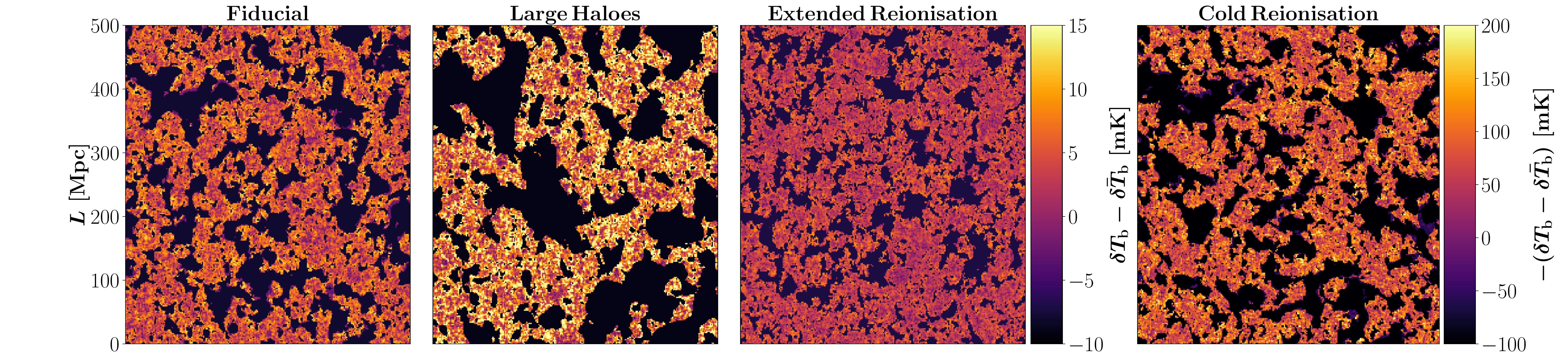}
		\includegraphics[trim = 3.3cm 0.6cm 0cm 0.5cm, scale = 0.288]{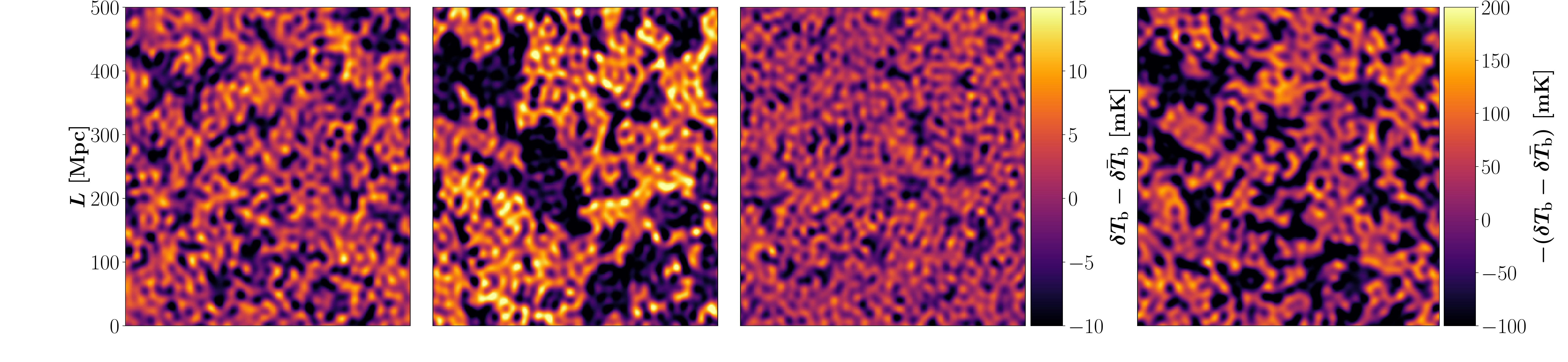}
		\includegraphics[trim = 3.3cm 0cm 0cm 0.5cm, scale = 0.288]{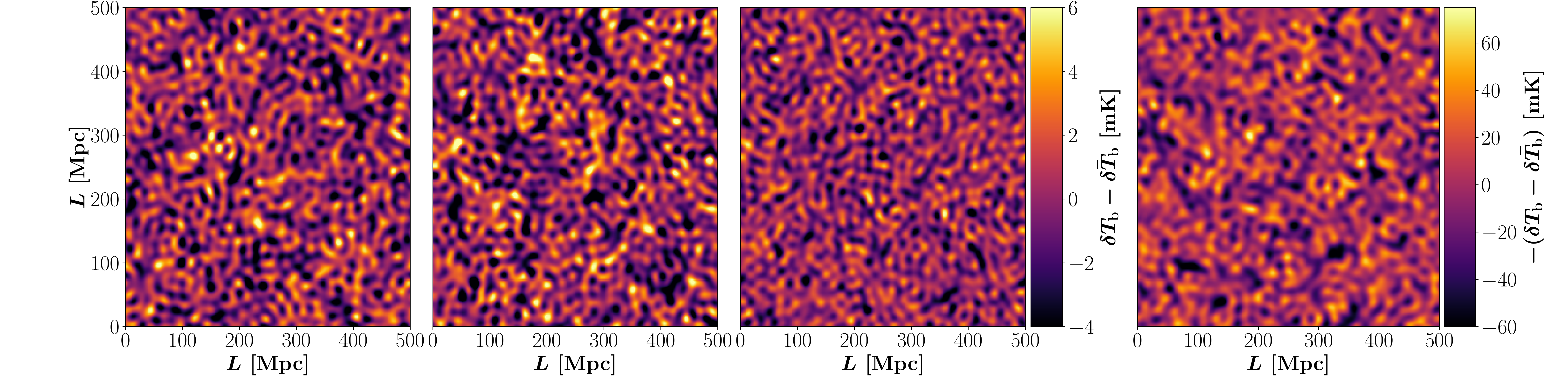}
		\includegraphics[trim = 3.3cm 0.8cm 0cm 0.2cm, scale = 0.395]{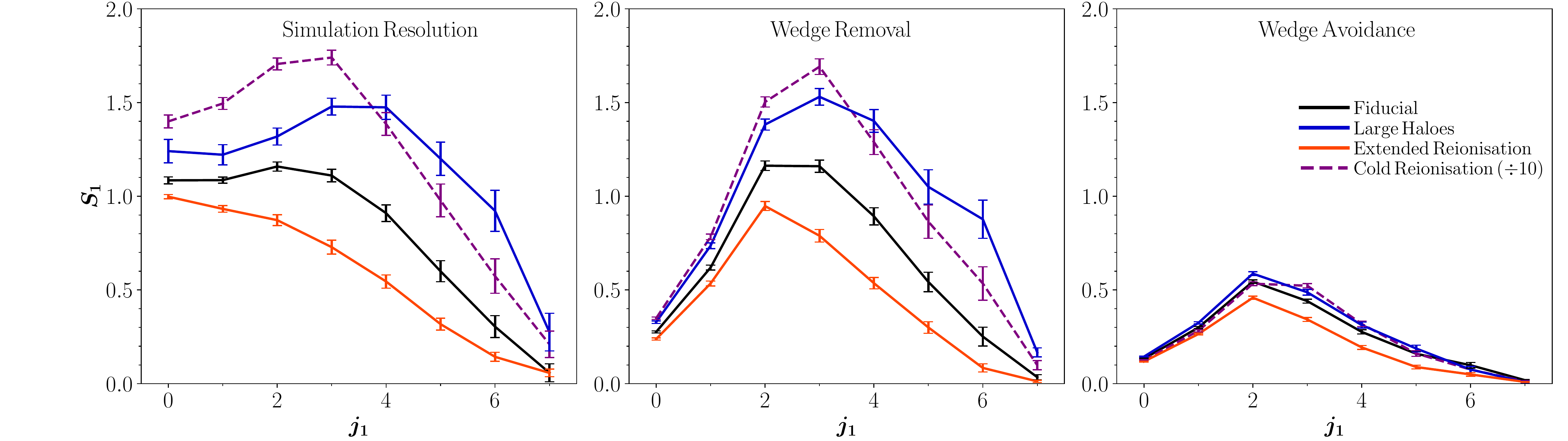}
	\end{center}
\caption[]{A visual comparison of the 21-cm signal at fixed IGM neutral fraction (\avenf~$\sim0.5$) for our four astrophysical models and in the presence of instrumental effects. For all, we present the 21-cm fluctuations (i.e. mean removed). \textit{Top row:} raw simulated output, \textit{Second row:} perfect foreground removal and \textit{Third row:} foreground avoidance. Note, for cold reionisation we multiply the signal by minus one to retain similar colour gradients (mean signal is negative). \textit{Bottom row:} The recovered first-order ($S_{1}$) WST scattering coefficients for each astrophysical model and raw simulation output, wedge removed and wedge avoidance, respectively. Error bars correspond to the 68th percentile uncertainty obtained from 50 different realisations (cosmic variance). Note, the amplitude of the cold reionisation coefficients have been reduced by a factor of 10 to be comparable with the other models.}
\label{fig:Instrument}
\end{figure*}

Unlike our earlier work we consider larger 21-cm simulations with transverse scales of 500 Mpc and 256 voxels per side length to more closely match the expected field-of-view of the SKA (see Section~\ref{sec:setup}). In the top row of Figure~\ref{fig:Instrument} we present a 2D snapshot of each astrophysical model at similar IGM neutral fractions (\avenf~$\sim0.5$) to more easily highlight the differences in each model. We also subtract the mean brightness temperature in each model to produce a zero mean signal, reminiscent of what is measured using radio interferometry.

\subsection{Instrumental effects}

Here, we outline our method for constructing realistic mock 2D images of the 21-cm signal using the SKA1--low, including the impact of thermal noise and astrophysical foregrounds. Since the visibilities ($uv$ coverage) of an interferometer experiment are frequency dependent, line-of-sight (frequency dependent) power can leak into transverse (frequency independent) Fourier modes. This results in a well-defined contaminated `wedge' in cylindrical 2D Fourier space \citep{Datta:2010p2792,Vedantham:2012p2801,Morales:2012p2828,Parsons:2012p2833,Trott:2012p2834,Thyagarajan:2013p2851,Liu:2014p3465,Liu:2014p3466,Thyagarajan:2015p7294,Thyagarajan:2015p7298,Pober:2016p7301,Murray:2018}. In this work, we consider two possible treatments of the astrophysical foregrounds when constructing our mock 21-cm images: (i) foreground removal, where we assume all foregrounds can be perfectly removed allowing us to use all available cosmological information and (ii) foreground avoidance, we can only using the pristine cosmological information above the foreground wedge. These choices should bookend what is plausible in practice.

To simulate the instrumental response ($uv$ coverage) and corresponding thermal noise power in 2D for the SKA we use a modified version of the publicly available \textsc{\small Python} module \textsc{\small 21cmSense}\footnote{https://github.com/jpober/21cmSense} \citep{Pober:2013p41,Pober:2014p35}. We use the SKA antenna configuration as outlined in the SKA System Baseline Design document\footnote{http://astronomers.skatelescope.org/wp-content/uploads/2016/09/SKA-TEL-SKO-0000422\textunderscore 02\textunderscore SKA1\textunderscore LowConfigurationCoordinates-1.pdf} which consists of 224 37.5m antennae stations distributed within a 500m radius. Finally, we assume a single, six-hour track per night for a total observing time of 1000 hours.

\subsubsection{Foreground Removal} \label{sec:removal}

To achieve mock 21-cm images assuming perfect foreground removal we perform the following steps:
\begin{itemize}
\item 2D Fourier transform the input (simulated) image
\item filter our image using the gridded $uv$-visibilities from \textsc{\small 21cmSense}, where modes sampled with finite $uv$-coverage are multiplied by unity while all others are zeroed
\item thermal noise is added to each cell by randomly sampling from the thermal noise power provided by \textsc{\small 21cmSense}
\item inverse Fourier transform to obtain our mock image
\end{itemize}
The second row of Figure~\ref{fig:Instrument} demonstrates these mock images for each astrophysical model. Despite the significantly lower spatial resolution, the ionised regions in all cases can be visually identified. For the large halo and cold reionisation models the features are much easier to identify due to their larger amplitude fluctuations in the 21-cm signal. Contrast this with the extended reionisation model, with considerably lower amplitude fluctuations.

\subsubsection{Foreground Avoidance}

To mimic foreground avoidance, contaminated modes located within the `wedge' in 2D cylindrical space given by
\begin{eqnarray} \label{eq:wedge}
k_{\parallel} =  mk_{\perp} + b
\end{eqnarray} 
are removed. Here $k_{\parallel}$ and $k_{\perp}$ are the line-of-sight and transverse Fourier modes, $b$ is a additive buffer of $\Delta k_{\parallel} = 0.1 \,h$~Mpc$^{-1}$ extending beyond the horizon limit and,
\begin{eqnarray}
m = \frac{D_{\rm C}H_{0}E(z){\rm sin}(\theta)}{c(1+z)},
\end{eqnarray} 
where $D_{\rm C}$ is the comoving distance, $H_{0}$ is the Hubble constant, $E(z) = \sqrt{\Omega_{\rm m}(1+z)^{3} + \Omega_{\Lambda}}$ and ${\rm sin}(\theta)$ is the telescope viewing angle, assumed conservatively to be $\theta = \pi/2$ (i.e. zenith pointing).

Our mock 21-cm images are then constructed by:
\begin{itemize}
\item constructing a 3D volume of the 21-cm signal centred on the desired observing frequency
\item 3D Fourier transforming this volume, filtering and adding thermal noise following the second and third steps in Section~\ref{sec:removal}
\item excise (zero) all modes located within the wedge
\item inverse Fourier transform back
\end{itemize}
Images in the third row of Figure~\ref{fig:Instrument} demonstrate this approach. Following foreground avoidance, it is considerably more difficult to visually identify the features in the 21-cm signal. However, as demonstrated in \citet{Greig:WST}, the WST is still capable of extracting astrophysical information.

\section{The Wavelet Scattering Transform} \label{sec:Understanding}

The primary goal of this work is to introduce a novel method for extracting the non-Gaussian information from images of the 21-cm signal using the WST. However, prior to this we first provide a summary of the WST. For more in-depth discussions on the WST and its interpretation in the context of the EoR, we defer the interested reader to \citet{Greig:WST}.

\subsection{Summary of the WST}

The WST is the convolution of an input image, $I(\boldsymbol{x})$, by a family of rotated and dilated wavelet filters (in this case Morlet filters). Each filter is defined by a physical scale, $j$, and rotation $l$. These physical scales correspond to a dyadic sequence ($2^{j}$) up to a maximum $J$ such that $2^{J}$ does not exceed the number of pixels in the image while the rotation angles are sampled at each $\pi/L$. After each convolution operation, the modulus of the filtered image is taken which in effect redistributes the information. This ensures that repeated filtering and modulus operations on the input image can then access higher-order spatial information. For example, filtering and taking the modulus of our input image a second time yields a measure of the strength of the clustering of spatial features (i.e. non-Gaussian information). In principle, this can be extended up to any arbitrarily large number of operations, though in practice it is rare to yield any useful information beyond the second-order. Finally, the filtered input image is spatially averaged to compress the available information into a single number, referred to as a scattering coefficient. Up to second order, these correspond to:
\begin{flalign}
s_{0} &= \langle I_{0}(x,y)\rangle, \\
s^{\,j_1,l_1}_{1} &= \langle \, | I_{0} \ast \psi^{j_1,l_1} \, | \rangle, \\
s^{\,j_1,l_1,j_2,l_2}_{2} &= \langle \, | I_{1} \ast \psi^{j_2,l_2} \, | \rangle = \langle \, | \, | I_{0} \ast \psi^{j_1,l_1} \,| \ast \psi^{j_1,l_1} \,| \ast \psi^{j_2,l_2} \, | \rangle. \label{eq:S2}
\end{flalign}

To reduce the total number of scattering coefficients to a more manageable number, we can average over all possible filter rotations as the cosmological signal does not have a preferred transverse direction,
\begin{flalign} 
S_{0} &= s_{0}, \\
S^{\,j_1}_{1} &= \langle s^{\,j_1,l_1}_{1} \rangle_{l_1}, \\
S^{\,j_1,j_2}_{2} &= \langle s^{\,j_1,l_1,j_2,l_2}_{2} \rangle_{l_1,l_2},
\end{flalign}
which results in a total of $1+J+J^{2}$ scattering coefficients for a given 21-cm image. In \citet{Greig:WST} we explored the redshift evolution of these scattering coefficients for four astrophysical models, in particular focussing on the similarities between the first-order scattering coefficients, $S_{1}$, and the 21-cm PS. In summary, the $S_{1}$ coefficients can be loosely interpreted as coarsely binned power spectra.

In the final row of Figure~\ref{fig:Instrument} we plot the $S_{1}$ coefficients as a function of increasing physical scale, $j$, for each astrophysical model directly calculated from the raw simulation output (left), after foreground removal (middle) or foreground avoidance (right). For our simulation setup, the spatial scales are logarithmically binned ranging from $\sim 5.2$~Mpc ($j=0$) to $\sim 666.6$~Mpc ($j=7$). Note here, we divide the amplitude of the cold reionisation $S_{1}$ coefficients by a factor of 10 in order to more clearly demonstrate the variation between astrophysical models. The error bars correspond to the estimated 68th percentile cosmic variance uncertainty obtained from 50 independent realisations.

As already discussed, the behaviour for the $S_{1}$ coefficients reflects that of the 21-cm PS, demonstrating a prominent bump characteristic of the maximum size of the ionised regions \citep[e.g.][]{Alvarez:2012p1930,Mesinger:2012p1131,Greig:2015p3675}. This feature occurs at larger scales for the large halo model due to the notably larger ionised regions down to small scales for the extended reionisation model (small ionised regions). The fiducial and cold reionisation models have bumps at equivalent scales as they have very similar reionisation morphologies, differing only in the amplitude of the temperature fluctuations due to different X-ray heating. The amplitude of these $S_{1}$ coefficients are proportional to the amplitude of the brightness temperature fluctuations.

The middle panel highlights the impact of thermal noise and finite instrumental resolution. Due to these effects, we lose information on the smallest spatial scales (below the instrument resolution) indicated by the drop in amplitude. However, on moderate to large scales, these remain relatively unaffected, and thus importantly remain sensitive to the astrophysical information. In the right most panel, we demonstrate the impact of foreground avoidance. This affects all scales, notably reducing the overall amplitude of the signal. Additionally, it makes it more difficult to isolate the bump corresponding to the sizes of ionised regions. Primarily this is due to the loss of contaminated foreground modes which impacts our ability to clearly identify the spatial extent of the ionised regions. However, the general shape still resembles that of the left most panel and as shown in \citet{Greig:WST}, foreground avoidance images still contain considerable astrophysical information that the WST is able to extract.

\subsection{Isolating Non-Gaussianity with the WST} \label{sec:isolate}

Wavelets inherently preserve the locality of the features within the input image. This property enables the non-Gaussian information to be easily extracted as each filtering operation simply measures the spatial clustering of the features present within the image (raw or filtered). That is, it utilises the important phase information for measuring the non-Gaussianity. This notably differs from the PS, whereby the phase information is destroyed and thus it only measures the amplitude of the spatial features (i.e. Gaussian information). Therefore, if we take our input image and randomise the phase information, we preserve the Gaussian (or $S_{1}$) information and destroy the non-Gaussian ($S_{2}$) information. In Appendix C of \citet{Greig:WST} we demonstrated this for the non-Gaussian information. Importantly, if the $S_{2}$ coefficients were only sensitive to non-Gaussian information then randomising the phase information would produce zero amplitude $S_{2}$ coefficients (i.e. no non-Gaussian information). However, the $S_{2}$ coefficients also depend on the $S_{1}$ coefficients (due to the repeated filtering, see Equation~\ref{eq:S2}) and thus result in reduced, but non-zero $S_{2}$ amplitude coefficients.

In this work, we take advantage of this feature in order to isolate out the purely non-Gaussian information. Our approach is as follows:
\begin{itemize}
\item[1.] from the input image, apply the WST to extract all second-order ($S_{2}$) scattering coefficients
\item[2.] randomise the phase information in the input image and extract the $S_{2}$ scattering coefficients from the randomised signal
\item[3.] repeat step two a large number of times (1000) to account for statistical uncertainties in the extracted scattering coefficients after phase randomisation
\item[4.] divide the $S_{2}$ coefficients from step one (cosmological signal) by those from step three (phase randomised). \item[5.] Any excess signal above unity is then indicative of non-Gaussian information within the 21-cm signal.
\end{itemize}
Note, this approach does not increase the sensitivity of the non-Gaussian signal, instead it simply makes it easier and cleaner to interpret. In \citet{Greig:WST} we explored the non-Gaussian information by considering the de-correlated $S_{2}$ coefficients, whereby the $S_{2}$ coefficients are divided through by the correlated $S_{1}$ information to separate out the non-Gaussian information. This results in coefficients of arbitrary amplitude. Instead, this approach represents the non-Gaussianity as an easy to interpret excess above unity.

\begin{figure*} 
	\begin{center}
		\includegraphics[trim = 2.1cm 0.3cm 0cm 0.9cm, scale = 0.38]{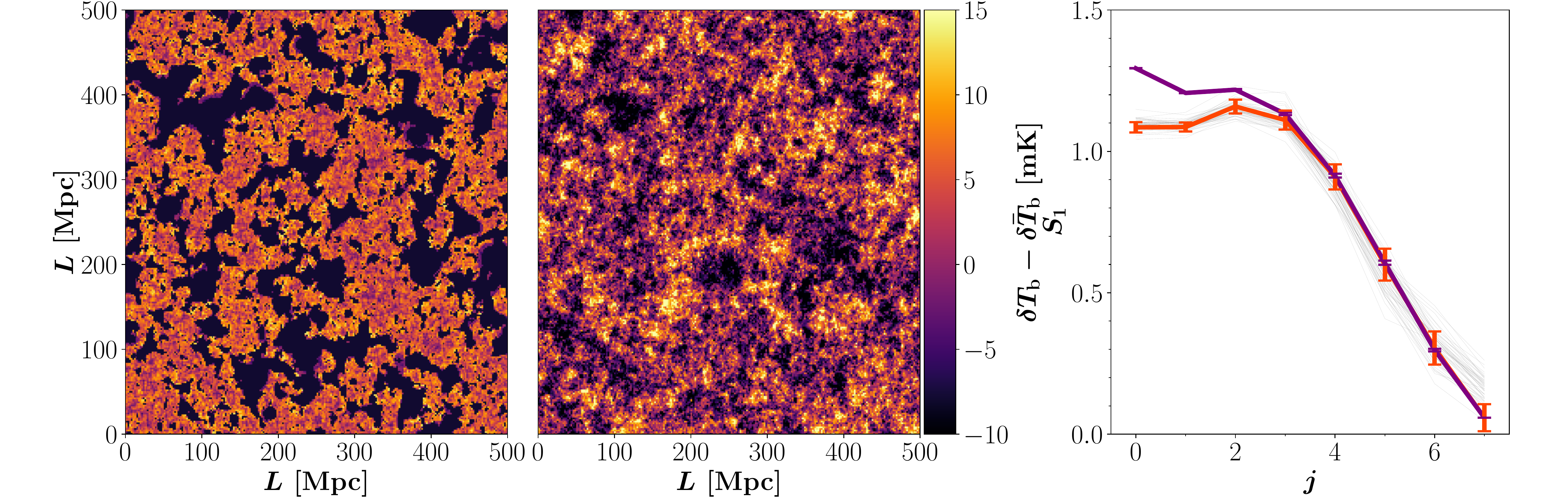}
		\includegraphics[trim = 1.7cm 0.6cm 0cm -0.1cm, scale = 0.563]{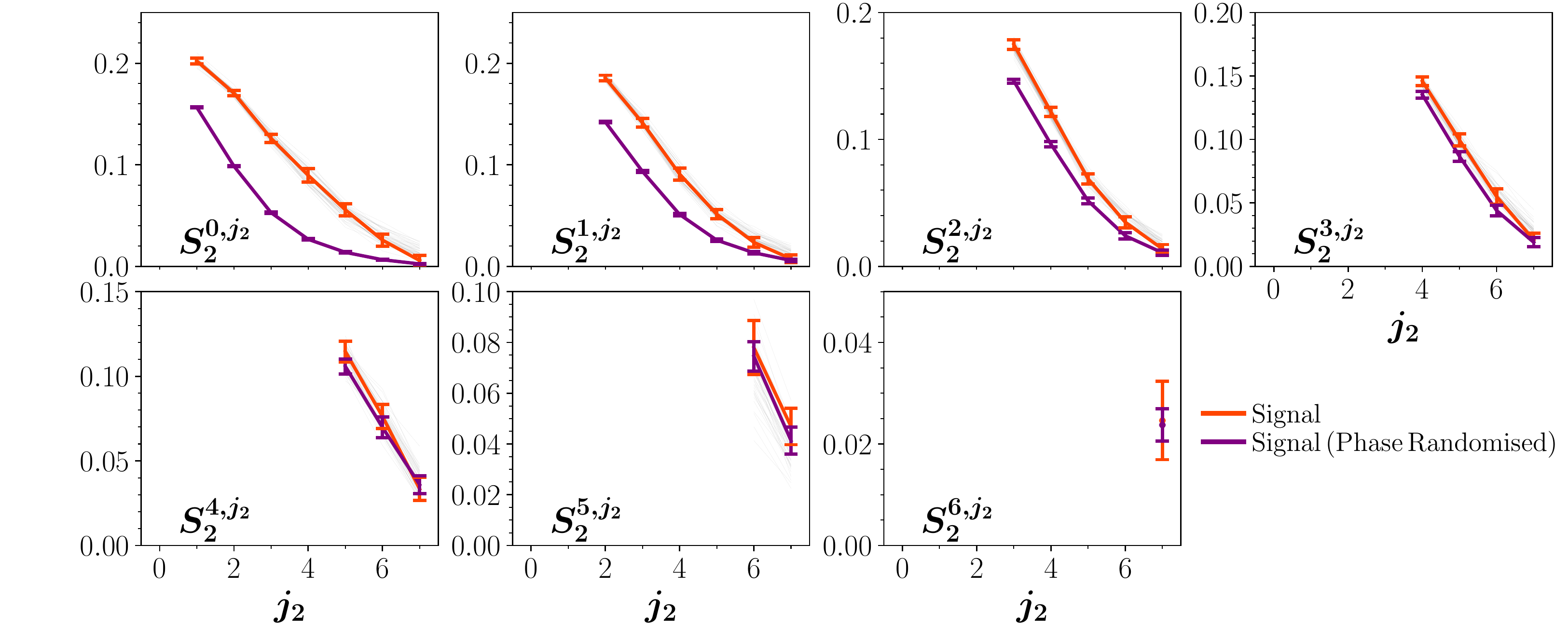}		
	\end{center}
\caption[]{Demonstrating phase randomisation of the input simulated 21-cm image for isolating non-Gaussian information with the WST. \textit{Top left:} The mean subtracted 21-cm signal from our fiducial simulation at $164$~MHz ($z\sim7.65$). \textit{Top middle:} The same image, but with randomised phase information. \textit{Top right:} The first-order ($S_{1}$) scattering coefficients from the true 21-cm signal (red curve) and randomised phase (purple). Grey curves correspond to 50 different cosmological realisations from which we estimate the 68th percentile cosmic variance error given by the red error bars. Purple curves and error bars are the mean scattering coefficients determined from a single image (i.e. one realisation) after 1000 different phase randomisations and the 68th percentile statistical uncertainty. \textit{Bottom panels:} The second-order ($S_{2}$) coefficients.}
\label{fig:PhaseInfo}
\end{figure*}

In Figure~\ref{fig:PhaseInfo} we provide a demonstration of this approach for our fiducial model at \avenf~$\sim0.5$ corresponding to $164$~MHz ($z\sim7.65$). We take our simulated 21-cm signal from our fiducial model (top left) and randomise the phase information (top middle). For completeness, in the top right, we also present the $S_{1}$ scattering coefficients from the true signal (red) and after phase randomisation (purple) while middle and bottom rows present all $S_{2}$ scattering coefficients (non-Gaussian information). The thin grey curves correspond to the scattering coefficients from 50 random cosmological realisations and the red error bars correspond to the 68 percentile cosmic variance uncertainty estimated from these 50 realisations. The purple curves correspond to the mean scattering coefficients after performing phase randomisation 1000 times for a single image (the top left image) and the associated errors are the 68th percentile statistical uncertainty (i.e. not cosmic variance).

For the $S_{2}$ coefficients, we observe a clear excess in the amplitude from the 21-cm signal to that from the randomised phases. This excess represents the non-Gaussian information held by the 21-cm signal. Note, the top right panel demonstrates that the $S_{1}$ coefficients are identical before and after phase randomisation only for $j > 2$. However, we expected that phase randomisation should preserve the first-order information across all scales. Importantly, the difference at $j < 3$ can be explained by the total energy norm of all scattering coefficients. In randomising the phase information, we preserve the total PS (i.e. Gaussian information). However, the $S_{1}$ coefficients do not directly map one-to-one to the PS as the PS is the square of the intensity while $S_{1}$ are just the intensity. Thus, the total energy norm of all scattering coefficients ($S_{1}$, $S_{2}$ and residuals) must equal the power spectrum. By randomising the phases we reduce the sparsity of the image (decrease $S_{2}$) which must be compensated for by an increase in $S_{1}$ to preserve the total power.

\section{Results} \label{sec:results}

\subsection{Experimental setup} \label{sec:setup}

The SKA intends to provide tomographic images of the 21-cm signal using a deep survey with an observing time of 1000 hr spanning 100 square degrees \citep{Koopmans:2015}. This total sky area is expected to be obtained from five independent 20 square degree fields. In this work, our chosen 500 Mpc simulations at 150~MHz ($z\sim8.5$) correspond roughly to $\sim3$ degrees on a side. Thus, we can achieve a similar sky area using 10 independent realisations. For all intents and purposes our choice of 10 independent images is equivalent to the intended survey design (five fields). The only potential differences will occur on the largest spatial scales, which will be most heavily affected by statistical noise and thus not sensitive in any case.

Throughout this work, we will only consider two observational frequencies: one at 177~MHz ($z\sim7$) and a second at 150~MHz ($z\sim8.5$). These choices are fairly arbitrary, but for our fiducial model correspond to the latter stages of reionisation (\avenf~$\sim0.25$), where the non-Gaussianity continues to increase due to the overlap of ionised regions and decreasing neutral patches, and just prior to the midpoint of reionisation (\avenf~$\sim0.6$) where the non-Gaussianity is more moderate (see Figure~7 of \citealt{Greig:WST}). These serve to provide a simple demonstration of the approach developed in this work to isolate the non-Gaussian signal. In practice, one would apply this approach across the full frequency coverage of the SKA (50-250~MHz) at 0.1~MHz intervals to extract the signal as a function of redshift.

To explore the detectability of the non-Gaussian signal with the SKA we generate 250 independent realisations (different initial conditions) of the 21-cm signal. For each, we then extract four different light-cone realisations by considering different starting locations in our simulation boxes (each location separated by 125~Mpc). This yields a total of 1000 independent 21-cm light-cones. We repeat this for all four astrophysical models.

\subsection{Maximum verses mean}

While in Figure~\ref{fig:PhaseInfo} the excess signal owing to non-Gaussianity is quite apparent, once we explore realistic mock images from the SKA in the presence of foregrounds it can become much harder to discern. As our survey strategy is to obtain multiple (10) images of the 21-cm signal, when analysing the results across our full survey footprint we have two possible approaches: (i) we simply take the mean of the excess signal across all images or (ii) we take the maximum signal from any one of the individual images to boost our sensitivity.

This idea of measuring the maximum signal stems from CNNs (from where the idea of the WST originates), where max pooling is used within the network architecture. Within this step, only those filtered images that maximise the particular feature of interest (i.e. most sensitive to that particular filter response) are kept in order to train the network. Thus, when analysing our images for the non-Gaussian 21-cm signal we consider only those images that maximise the signal. In other words, we anticipate that the non-Gaussian information will be largest within one (or a few) outlier images from the full sample. Whereas if we take the mean, we will wash out the signal reducing our overall sensitivity. 

\begin{figure*} 
	\begin{center}
		\includegraphics[trim = 1.3cm 0.7cm 0cm 0.7cm, scale = 0.55]{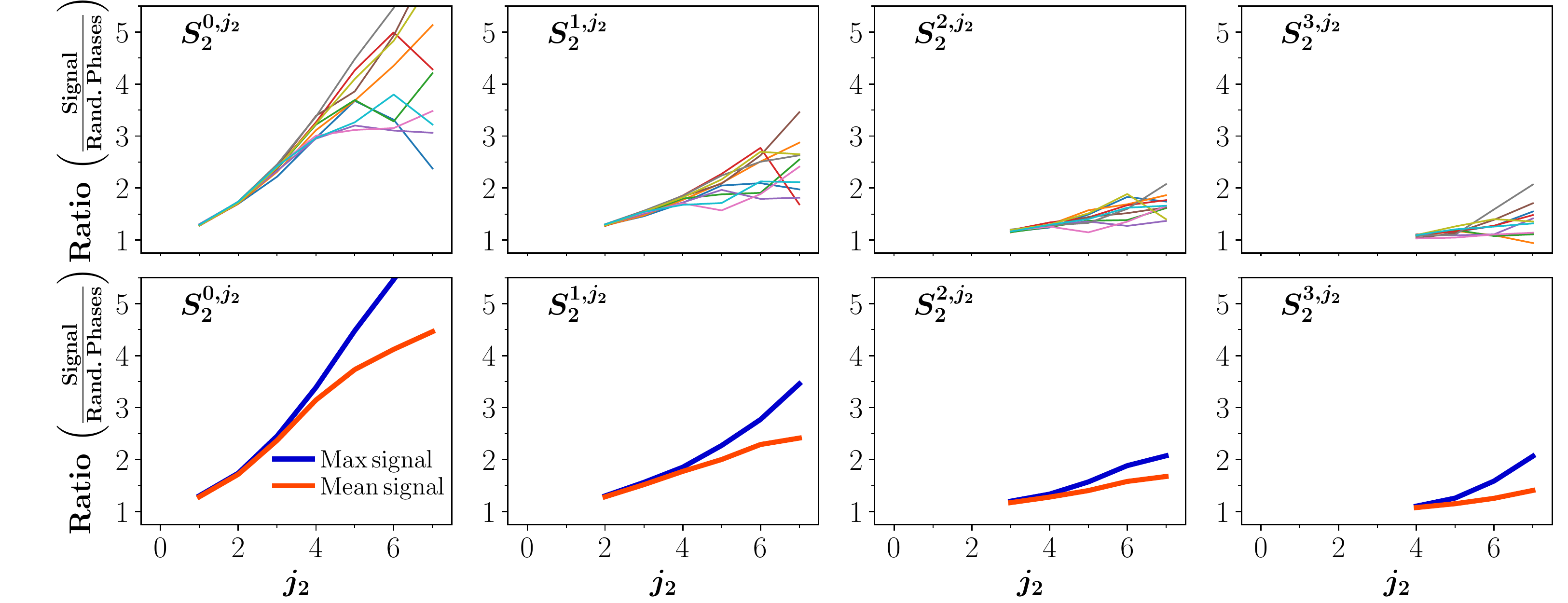}
	\end{center}
\caption[]{The ratio of the $S_{2}$ coefficients from an image of the 21-cm signal relative to those after phase randomisation at $164$~MHz ($z\sim7.65$). \textit{Top row:} the ratio from 10 different realisations of the 21-cm signal. \textit{Bottom row:} comparison of the recovered signal when considering the maximum (blue curve) or mean signal (red). See text for further details.}
\label{fig:RawRatio}
\end{figure*}

In Figure~\ref{fig:RawRatio} we demonstrate this approach using simulated images of the 21-cm signal with no instrumental or foreground effects at $164$~MHz ($z\sim7.65$). In the top row we present the ratio of the $S_{2}$ coefficients from the full signal to those after phase randomisation for 10 independent realisations. In the bottom row, we obtain the mean (red) and maximum (blue) signal across these images. In all cases, this ratio is in excess of unity, indicative of a strong detection of the non-Gaussian signal. Further, this maximum signal significantly boosts the amplitude of the signal over the mean, which will become important once instrumental effects are included. Since the scattering coefficients are correlated across different scales, in most cases this maximum signal originates from just one (or a few) similar images (i.e. the outliers of the distribution). Nevertheless, throughout our results below we will always present both the mean and the maximum non-Gaussian signal.

\subsection{Foreground removal} \label{sec:FR}

\begin{figure*} 
	\begin{center}
		\includegraphics[trim = 0.5cm 0.8cm 0cm 0.5cm, scale = 0.55]{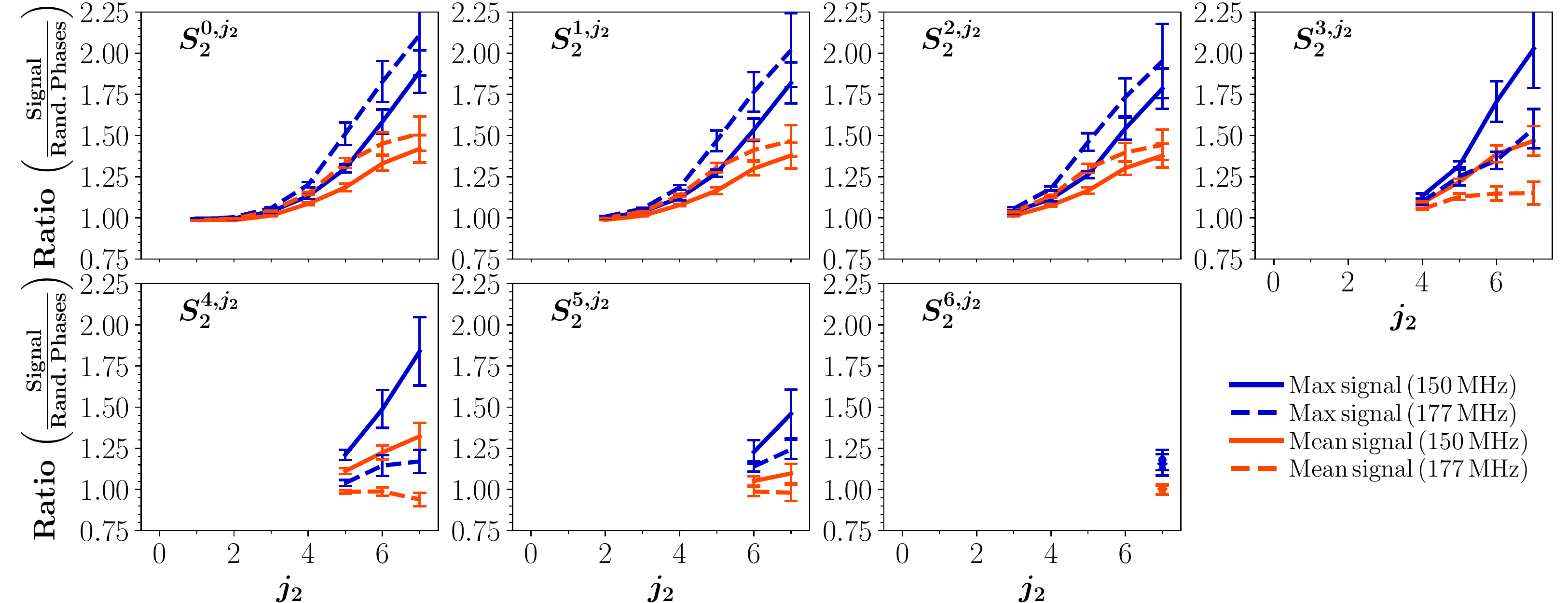}
	\end{center}
\caption[]{The ratio of second-order, $S_{2}$, scattering coefficients before and after phase randomisation assuming a 1000hr observation at 150 MHz ($z\sim8.5$; solid curves) and at 177 MHz ($z\sim7$; dashed curves) with the SKA and astrophysical foregrounds perfectly removed (i.e. no wedge). Blue (red) curves correspond to the mean of the maximum (mean) ratio after extracting 10 independent images of the 21-cm signal 1000 times. Error bars correspond to the 68th percentiles, which in effect includes cosmic variance as we sample across a large number of independent realisations.}
\label{fig:ratio_nowedge}
\end{figure*}

First, we consider the optimistic scenario whereby all foreground contamination can be removed from our 21-cm images. For this we only consider our fiducial model, but for completeness we will explore all four astrophysical models in Section~\ref{sec:diffmodels} assuming wedge avoidance. 

\subsubsection{Statistical detection of non-Gaussianity}

To explore this within a statistical framework, we perform the following:
\begin{itemize}
\item randomly select 10 mock 21-cm images to achieve the SKA survey footprint
\item isolate the non-Gaussian information following the steps outlined in Section~\ref{sec:isolate}
\item determine the mean and maximum ratio of non-Gaussianity across these 10 images
\item repeat the above steps 1000 times in order to yield a statistical distribution of the mean and maximum non-Gaussian signal
\end{itemize}

In Figure~\ref{fig:ratio_nowedge}, we present the ratio of the $S_{2}$ scattering coefficients before and after phase randomisation indicating the non-Gaussian information within the 21-cm signal. The curves correspond to the mean of our distributions obtain from our 1000 realisations for the mean (red) and maximum (blue) non-Gaussian signal. The solid (dashed) curves correspond to our different observing frequencies, 150 and 177~MHz, respectively. The error bars correspond to the 68th percentiles estimated from the distribution of realisations. In effect, these error bars take into account cosmic variance uncertainty as they are sampled over a distribution of randomly selected cosmological realisations.

For the first few $S_{2}$ coefficients, the amplitude of this ratio drops by a factor of $\sim3-4$ relative to the raw simulated images used in Figure~\ref{fig:RawRatio} due to the finite resolution of the SKA (see Figure~\ref{fig:Instrument}). For all larger scale $S_{2}$ coefficients, the amplitude of this ratio remains comparable to that from the raw simulated images indicating that the non-Gaussian 21-cm signal should be easily observable with the SKA, albeit assuming perfect foreground removal. 

This remains true irrespective of the observing frequency or whether using the mean or maximum amplitude $S_{2}$ coefficients. Nevertheless, we clearly see the advantage of using this maximum signal, with the amplitude of the ratio typically being a factor of two larger than the mean signal. However, as one might expect given that the maximum signal depends on outliers from the distribution, the statistical uncertainty on the maximum signal is larger compared to that of the mean signal. Importantly, this increase in the amplitude outweighs the larger statistical uncertainty which results in a stronger overall detection of the non-Gaussian signal (excess of the ratio above unity). To some extent, the maximum signal can be considered as an optimistic measurement whereas the mean signal is more conservative.

Note, for $j_{1}<3$ we observe the amplitude of the ratio to be larger at 177~MHz than at 150~MHz (i.e. more non-Gaussianity), however this swaps for $j_{1}\geq3$. This likely occurs due to differences in the ionisation morphology between these frequencies. At $j_{1}<3$ we are sensitive to the strength of the clustering of small-scale features, whereas at $j_{1}\geq3$ we are more sensitive to the clustering strength of intermediate to large-scales. At 177~MHz (\avenf~$\sim0.25$), the IGM contains many more smaller, isolated neutral patches increasing the amplitude of the non-Gaussianity. At 150~MHz (\avenf~$\sim0.6$), the neutral patches are larger and more connected resulting in a higher amplitude signal on these intermediate to large scales than at 177~MHz.

\subsubsection{Non-Gaussianity signal-to-noise}

Finally, to more clearly quantify the detection of the non-Gaussian signal, we define a signal-to-noise (S/N). This is defined for a single image as,
\begin{eqnarray} \label{eq:StoN}
{\rm S/N} = \frac{S^{j_{1},j_{2}}_{2, {\rm cosmological}} - \bar{S}^{j_{1},j_{2}}_{2, {\rm phase\,randomised}}}{\sigma_{S^{j_{1},j_{2}}_{2, {\rm phase\,randomised}}}},
\end{eqnarray}
where $S^{j_{1},j_{2}}_{2, {\rm cosmological}}$ is the second-order scattering coefficient from the observed 21-cm image and $\bar{S}^{j_{1},j_{2}}_{2, {\rm phase\,randomised}}$ and $\sigma_{S^{j_{1},j_{2}}_{2, {\rm phase\,randomised}}}$ are the mean and standard deviation of the second-order scattering coefficients obtained after randomising the phases 1000 times (i.e. no non-Gaussian information). Note, as this S/N is defined for a single image this uncertainty is purely the statistical uncertainty. In effect, this quantity represents the number of standard deviations (statistical) that the excess amplitude sits above unity (the non-Gaussian information).

\begin{figure*} 
	\begin{center}
		\includegraphics[trim = 1.3cm 0.8cm 0cm 0.5cm, scale = 0.56]{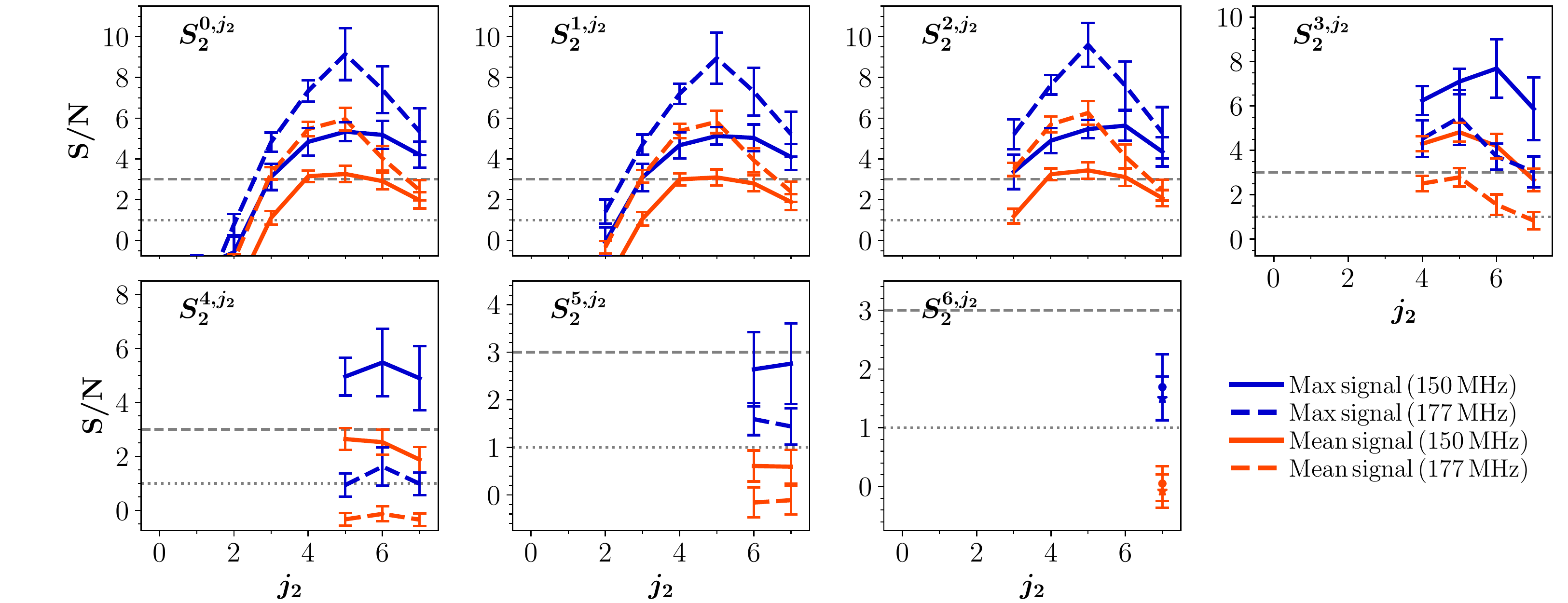}
	\end{center}
\caption[]{The corresponding signal-to-noise of the non-Gaussian signal for the second-order scattering coefficients assuming a 1000hr observation at 150 MHz ($z\sim8.5$; solid curves) and at 177 MHz ($z\sim7$; dashed curves) with the SKA and astrophysical foregrounds removed (i.e. no wedge). Blue (red) curves correspond to the maximum (mean) ratio after extracting 10 independent images of the 21-cm signal 1000 times. Error bars correspond to the 68th percentiles which in effect include cosmic variance as we sample across a large number of independent realisations.}
\label{fig:StoN_nowedge}
\end{figure*}

In Figure~\ref{fig:StoN_nowedge} we present the mean S/N assuming perfect foreground removal, performing the same statistical analysis as previously (randomly selecting 10 images 1000 times). To more clearly indicate detection limits, the horizontal grey dashed (dotted) curves correspond to a mean S/N of three (one). The solid (dashed) curves correspond to the mean statistical S/N determined from 1000 realisations of our 10 image survey while the error bars are the 68th percentile uncertainty. As we vary over a large number of independent realisations, this uncertainty accounts for cosmic variance. Assuming perfect foreground removal we can recover relatively strong detections of the non-Gaussian signal with S/N~$\gtrsim8$ for the maximum signal, decreasing down to S/N~$\gtrsim5$ for the mean signal at 177~MHz. This then decreases to a more modest S/N~$\gtrsim5$ (3) at 150~MHz for the maximum (mean) signal. Overall, we consistently find that using the maximum signal increases the S/N on all scales by $\Delta$(S/N)~$=2-3$. As before, the S/N of the non-Gaussianity is largest at $j_{1}<3$ at 177~MHz, switching to a stronger detection at $j_{1}\geq3$ at 150~MHz. For the largest scales, $j_{1} > 4$ the non-Gaussian signal drops away fairly rapidly, indicating very little sensitivity on these spatial scales ($\gtrsim167$~Mpc). Note, here and throughout this work we always assume the full 1000~hr observing time with the SKA. In Appendix~\ref{sec:obs_time} we explore the impact of shorter observing times.

Importantly, the strength of the non-Gaussian signal reported here is likely not the maximum achievable value. While we have selected two observing frequencies for our fiducial model where we know there should be non-Gaussian signal from \citet{Greig:WST}, these were not where it was maximal. Further, the performance of the SKA (both in terms of resolution and noise properties) improves to higher frequencies (lower redshifts), corresponding to the latter stages of reionisation (where non-Gaussianity also increases). Finally, as we shall see in Section~\ref{sec:diffmodels} alternative astrophysical models for the EoR can yield stronger reported detections. However, the main purpose of this work was to demonstrate the method to isolate the non-Gaussian signal along with an example, not to highlight the largest possible detection of non-Gaussianity.

Expanding this approach over the entire SKA observable bandwidth, one can detect the S/N of the non-Gaussian signal as a function of redshift. Doing so, would enable us to discriminate various astrophysical model parameters responsible for the EoR. However, as this was the primary focus of \citet{Greig:WST} we do not explore this further here.

\subsection{Foreground avoidance} \label{sec:FA}

\subsubsection{Fiducial model}

\begin{figure*} 
	\begin{center}
		\includegraphics[trim = 0.5cm 0.3cm 0cm 0.5cm, scale = 0.55]{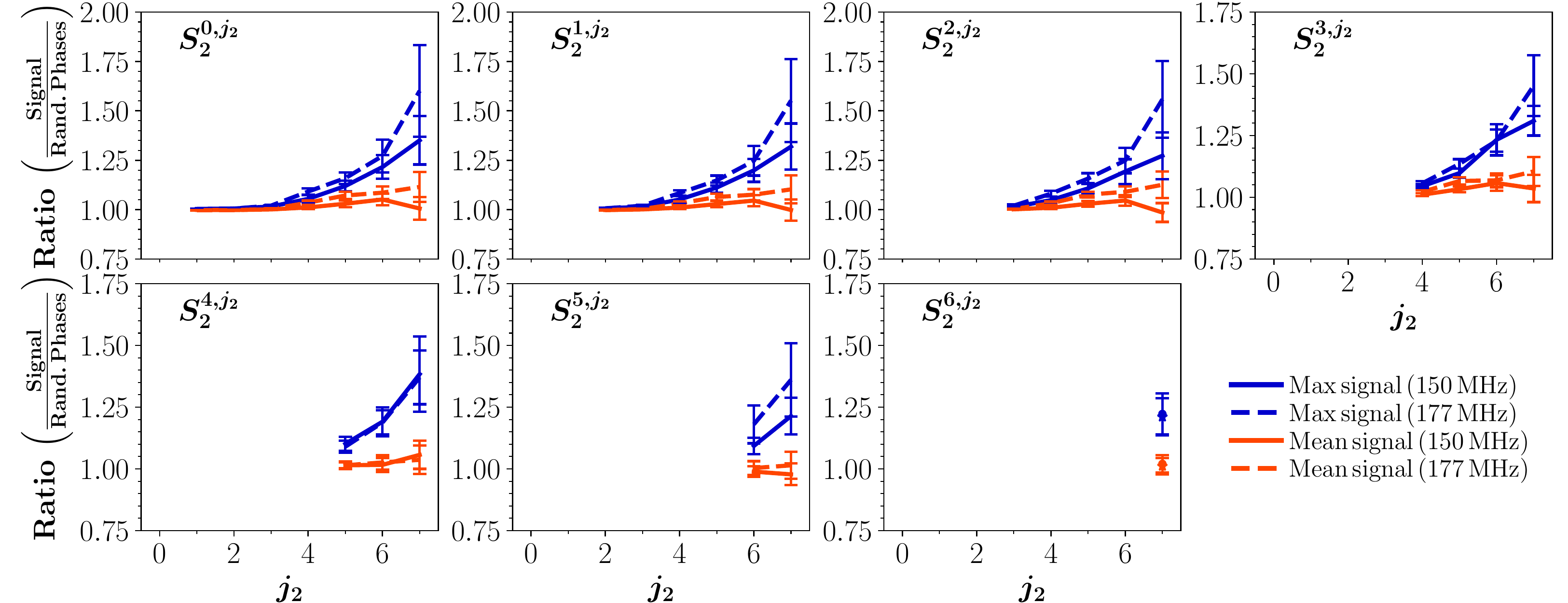}
	\end{center}
\caption[]{The same as Figure~\ref{fig:ratio_nowedge} except with the foreground contaminated (i.e. wedge) modes additionally removed.}
\label{fig:ratio_wedge}
\end{figure*}

\begin{figure*} 
	\begin{center}
		\includegraphics[trim = 1.3cm 0.3cm 0cm 0.5cm, scale = 0.56]{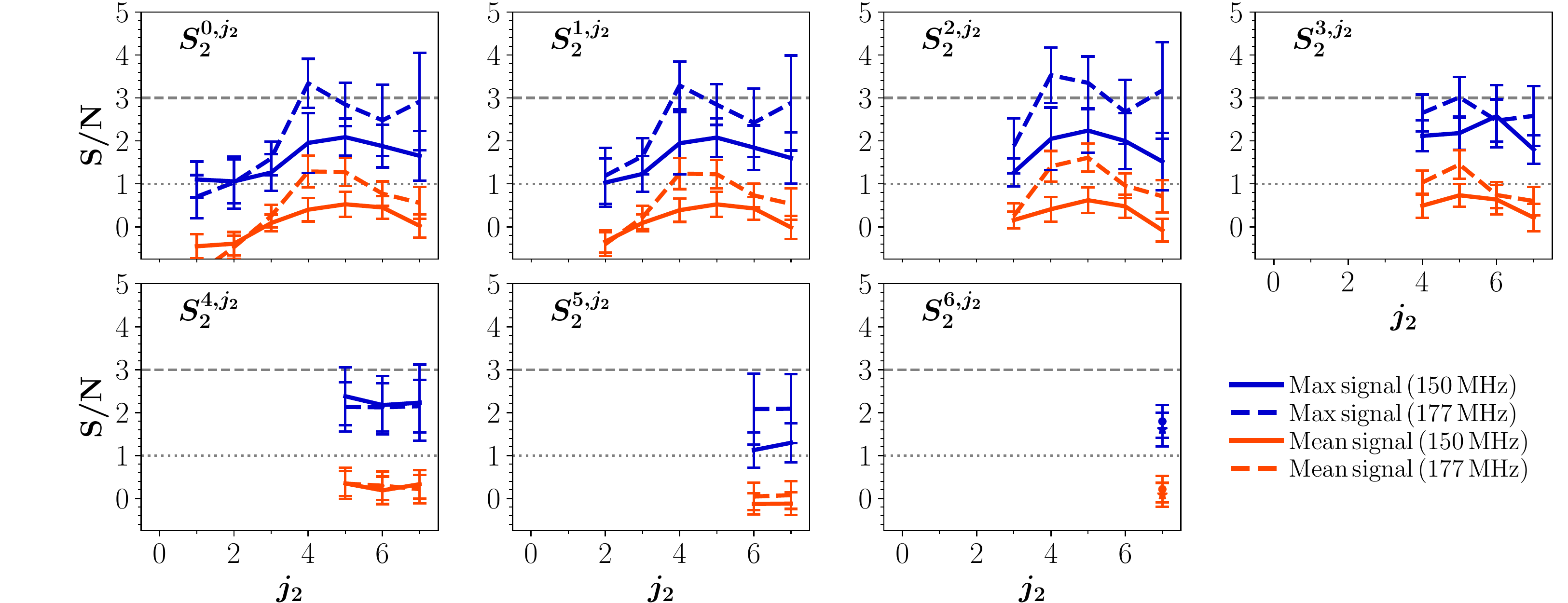}
	\end{center}
\caption[]{The same as Figure~\ref{fig:StoN_nowedge} except with the foreground contaminated (i.e. wedge) modes additionally removed.}
\label{fig:StoN_wedge}
\end{figure*}

Here, we repeat our previous analysis instead assuming foreground avoidance. In Figure~\ref{fig:ratio_wedge} we present the ratio of the $S_{2}$ coefficients while in Figure~\ref{fig:StoN_wedge} we present the resulting S/N. Firstly, removing wedge modes further decreases the amplitude of this ratio by an additional $20-60$~per cent. Further, this reduction is present across all scales rather than just the smallest scales which were lost due to the instrument resolution. This decrease can be attributed to the loss of the spatial information that resides in the contaminated wedge region. Indeed, for the mean signal (red curves) this ratio barely sits above unity, necessitating the need for our exploration of the maximum signal.

For our two observational frequencies, we observe a notable decrease in the overall S/N of the non-Gaussian signal, recovering at best a S/N~$\sim3$ at 177~MHz when the non-Gaussianity is considerably larger. At 150~MHz, we at best achieve a S/N~$\sim2$. Note, we do not observe the same flip between the two observing frequencies at $j_{1}=3$ as seen previously. This is because wedge avoidance more significantly impacts large scales relative to the small scales, affecting progressively larger $j_{1}$'s. Nevertheless, even after conservatively removing all contaminated foreground modes we can still recover a statistically significant detection of the non-Gaussianity in the 21-cm signal, provided we use the maximum signal.

Using the mean signal (red curves), only a very marginal detection (S/N~$\sim1$) would be expected, and only at 177~MHz. At 150~MHz, the mean signal barely sits above zero implying no detection would be possible. Thus, under foreground avoidance with only the mean signal, non-Gaussianity is likely only detectable near the final stages of reionisation when it is expected to be at its largest.

\subsubsection{Different astrophysical models} \label{sec:diffmodels}

Having thus far only considered our fiducial model, we now explore the detectability of the non-Gaussianity across our four astrophysical models. Here, we only consider the conservative case of foreground avoidance, but expect the overall trends to be similar for perfect foreground removal except at a higher S/N (nominally an increase of $\Delta$(S/N)~$=2-4$).

\begin{figure*} 
	\begin{center}
		\includegraphics[trim = 0.5cm 0.3cm 0cm 0.5cm, scale = 0.55]{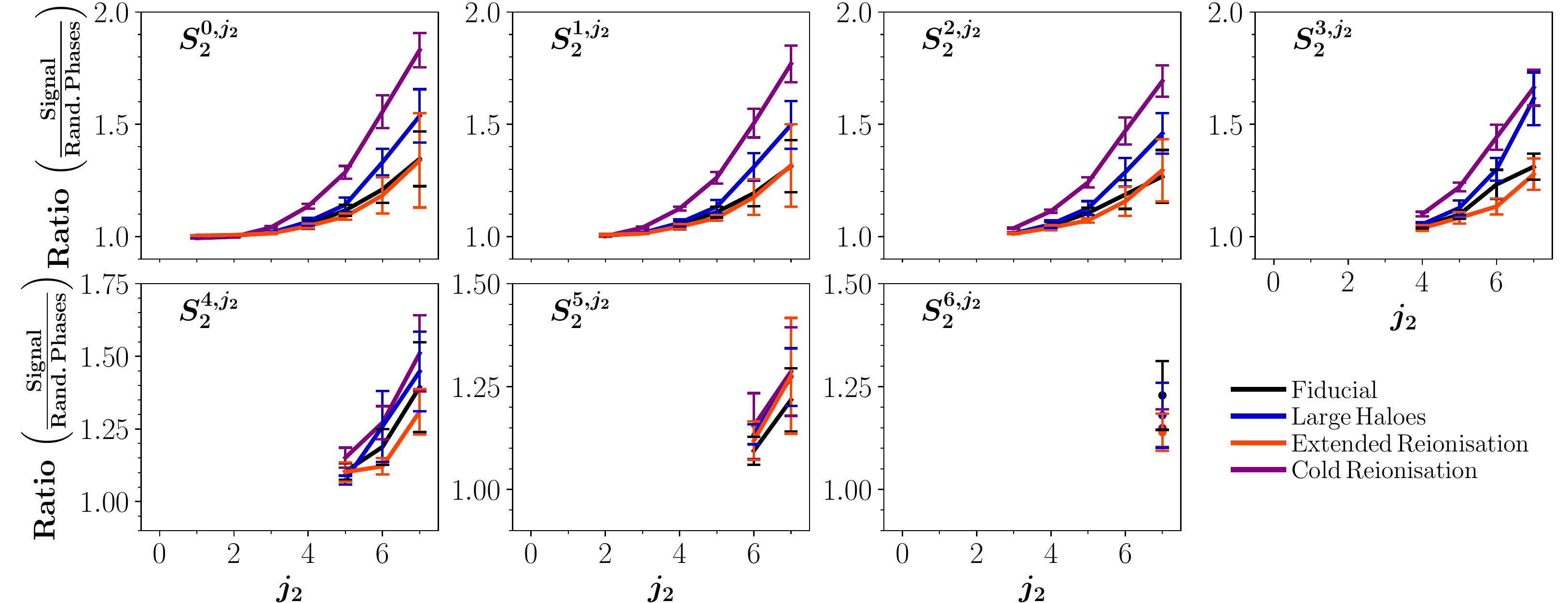}
		\includegraphics[trim = 0.5cm 0.7cm 0cm -0.3cm, scale = 0.55]{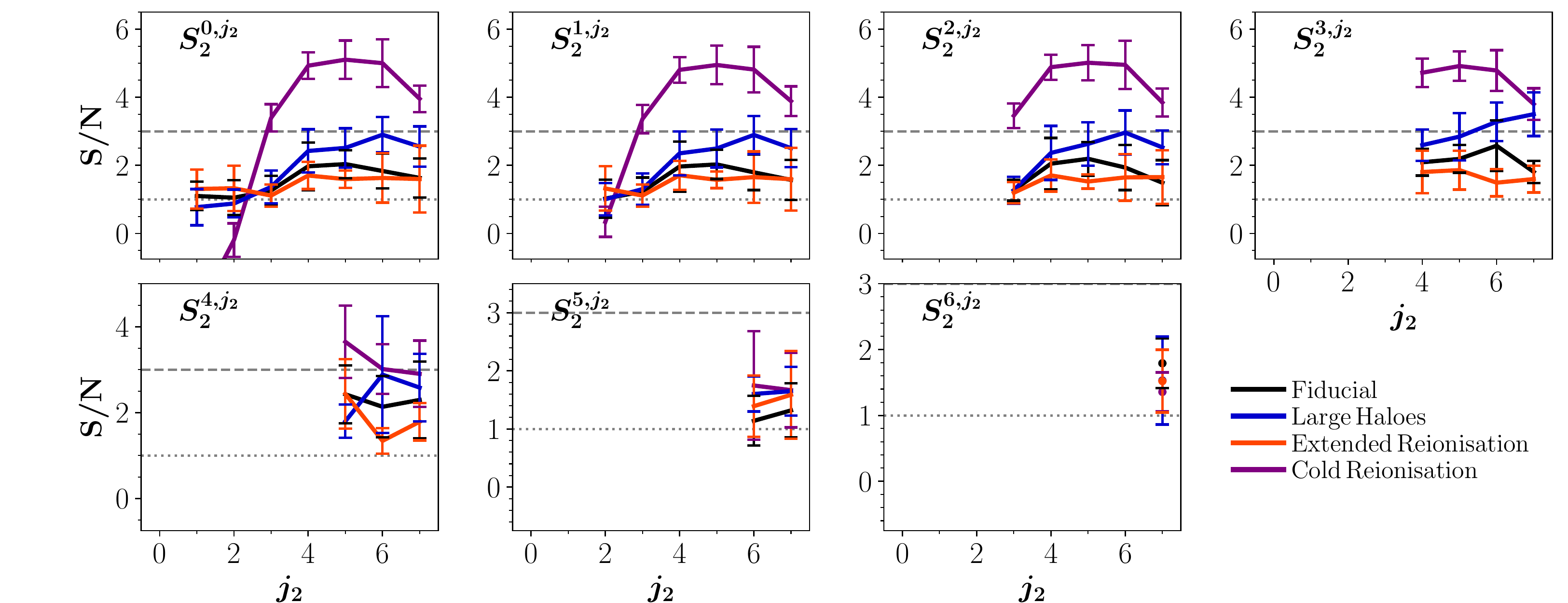}
	\end{center}
\caption[]{The ratio (top two rows) and corresponding signal-to-noise (bottom two rows) of the second-order, $S_{2}$, scattering coefficients following phase randomisation for our four astrophysical models: fiducial (black curves), large haloes (blue), extended reionisation (red) and cold reionisation (purple). Here, we assume a 1000hr observation at 150 MHz ($z\sim8.5$) with foreground avoidance and the curves correspond to the maximum signal extracted from 10 independent images of the 21-cm signal. Error bars are the 68th percentiles extracted from 1000 randomly drawn samples of 10 independent images and thus take into account cosmic variance.}
\label{fig:WA_150}
\end{figure*}

In the top half of Figure~\ref{fig:WA_150} we present the ratio of the $S_{2}$ coefficients following phase randomisation at 150~MHz ($z\sim8.5$) while in the bottom half we present the detectability of the signal with the S/N. For the cold reionisation (purple) and large haloes (blue) models we clearly see a much larger amplitude signal, owing to the increased contrast in the brightness temperature fluctuations (see Figure~\ref{fig:Instrument}). With respect to the S/N this translates to a S/N~$\sim5$ for most scales for cold reionisation but only S/N~$\sim2.5-3$ for large haloes (slightly higher than the S/N~$\sim2$ for our fiducial model and the extended reionisation model). However, note that this is at a fixed frequency, thus each model is at a different stage of the reionisation process (see Figure~2 of \citealt{Greig:WST}). For example, the large halo model is in the earlier stage of reionisation (\avenf~$\sim0.75$) in contrast to all others which are at (\avenf~$\sim0.6$). Thus, we would expect a stronger detection for the large halo model at a comparable stage of reionisation history.

While it is plausible to distinguish some of these models at a single frequency (e.g. the cold reionisation and large halo model from our fiducial model), it would be possible to tweak the astrophysical parameters making them indistinguishable. Thus, for robust model inference, one really requires detections across a broad range of frequencies (redshifts), which has been explored in \citet{Greig:WST}. Here, we are only interested in providing a demonstration of this novel approach to isolate the non-Gaussian signal.

\begin{figure*} 
	\begin{center}
		\includegraphics[trim = 0.5cm 0.7cm 0cm 0.5cm, scale = 0.55]{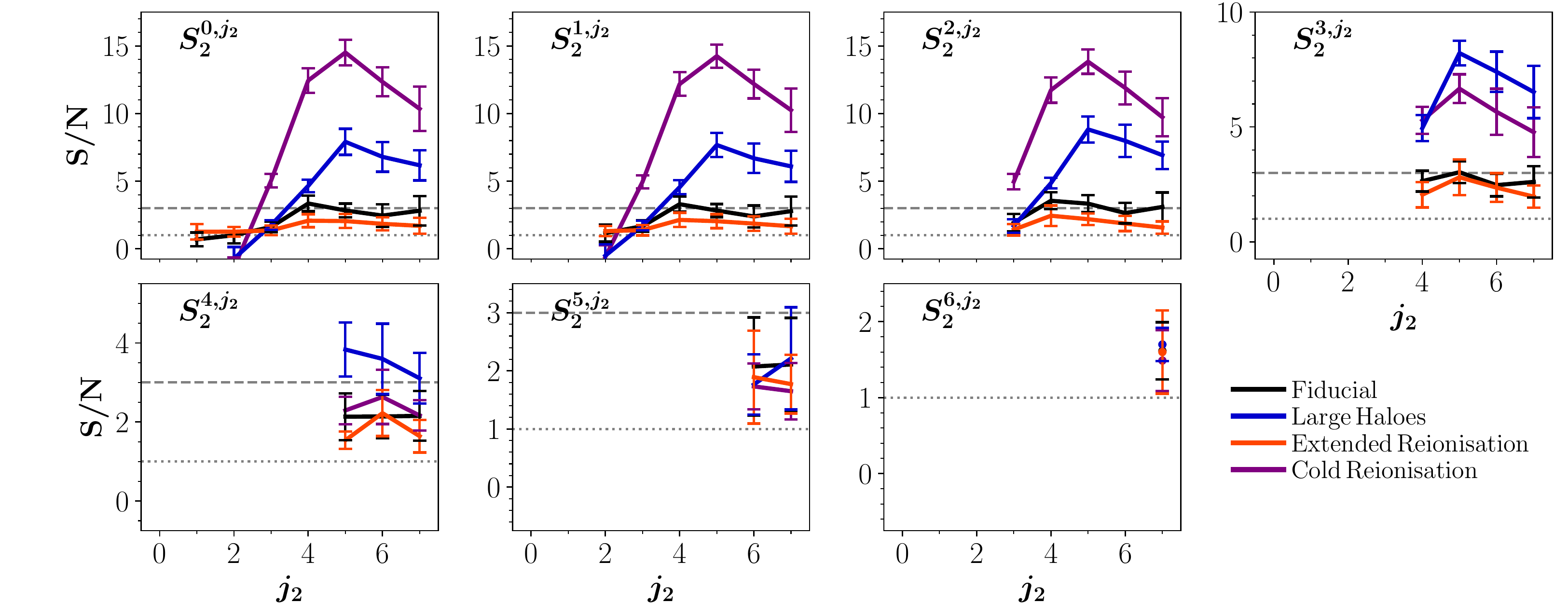}
	\end{center}
\caption[]{The same as Figure~\ref{fig:WA_150} except at 177 MHz ($z\sim7$).}
\label{fig:WA_177}
\end{figure*}

In Figure~\ref{fig:WA_177} we additionally present the S/N of the non-Gaussianity for our four astrophysical models at 177~MHz ($z\sim7$). Note, we do not present the ratio of the $S_{2}$ coefficients at this frequency as they are similar to those at 150~MHz, just at a larger amplitude. As expected, since the non-Gaussianity is larger towards the tail end of reionisation, we observe considerably larger S/N detections. For example, we obtain a S/N~$\sim14~(8)$ for our cold reionisation (large haloes), respectively. The extended reionisation model on the other hand returns a relatively weak detection (S/N~$\sim1-2$). However, this is effectively by design, in this model the IGM is still 40 per cent neutral. Thus, it is yet to reach the tail of reionisation whereby the non-Gaussianity grows significantly.

\section{Discussion} \label{sec:discussion}

Thus far, we have solely focussed on introducing our method to isolate the non-Gaussianity of the 21-cm signal along with  a simple example for two different observing frequencies and treatments of the astrophysical foregrounds. Here, we now provide a brief discussion of several caveats and assumptions within this work.

Firstly, when estimating the amplitude of the thermal noise to add to our mock images, we use the standard expression to describe the sky temperature of $T_{\rm sky} = 60\left(\frac{\nu}{300~{\rm MHz}}\right)^{-2.55}~{\rm K}$ \citep{Thompson2007}. However, an alternative estimate of the sky temperature \citep{DeBoer:2017p6740,Munoz:2022} yields a factor of three lower estimate for the sky temperature. With this latter expression, the overall amplitude of the noise fluctuations would decrease by an equivalent amount, increasing our overall sensitivity to the 21-cm brightness temperature fluctuations. In fact, recent estimates from the MWA \citet{Rahimi:2021} tend to suggest lower sky temperatures that those predicted by the expression above. Thus, the recovered results within this work can be considered to be a conservative estimate.

Further, for our foreground avoidance model we have assumed an additive buffer ($b$ in Equation~\ref{eq:wedge}) above the foreground contaminated wedge. For this, we conservatively chose $\Delta k_{\parallel} = 0.1 \,h$~Mpc$^{-1}$. Removing this, or considering a smaller buffer region (i.e. assuming the foregrounds do not bleed out as far from the wedge) would increase the overall detectability of the non-Gaussian signal assuming foreground avoidance. Thus again, our results can be considered conservative due to this choice.

Importantly, throughout this work we have assumed a fairly simplistic treatment of the astrophysical foregrounds and associated systematics of a 21-cm signal detection. That is, assuming perfect removal (Section~\ref{sec:FR}) or that all foregrounds exist within the contaminated wedge region leaving a pristine window into the cosmological signal above it (foreground avoidance; Section~\ref{sec:FA}). However, in practice, there are many potential artefacts that would be expected to remain in any realistic 21-cm image. Further, these are likely to contribute non-trivially, contaminating the non-Gaussian cosmological signal. For example, residuals owing to the incomplete removal of foreground radio sources would yield a non-Gaussian signature in the resultant 21-cm images. In particular, this can occur when the ionosphere is active and any direction-dependent source peeling does not correctly capture the true source position. This leads to an over/under-subtraction of the sources (peaks and holes; \citealt[e.g.][]{Jordan:2017,Chege:2021}). Thus, the detection of any non-Gaussianity within a 21-cm image following the approach outlined in this work may not be entirely cosmological in origin. However, this likely can be mitigated to some extent by considering the redshift evolution of the non-Gaussian component as systematic artefacts should not evolve as a function of redshift (or will evolve in a manner distinguishable from the cosmological signal). Further, some second-order scattering coefficients may be more heavily affected than others by these artefacts (e.g. scale dependent systematics) leaving others relatively free of systematics. Thus it should be possible to separate the non-Gaussian cosmological signal from astrophysical systematics. However, this requires a more detailed exploration beyond the scope of this current work, thus we defer it to future work.

\section{Conclusion} \label{sec:conclusion}

We introduced the wavelet scattering transform (WST) for analysing 2D images of the 21-cm signal during reionisation in \citet{Greig:WST}. In particular, we demonstrated that the WST can outperform the standard 21-cm power spectrum in the context of astrophysical parameter inference. Primarily, this occurs due to the ease in which the WST can access the non-Gaussian information in the 21-cm signal by successively applying wavelet filters to analyse the spatial features in the input images.

In this work, we introduced a novel method to more easily isolate the non-Gaussian information from 2D images using the WST. This approach compares the second-order ($S_{2}$) scattering coefficients extracted from the original cosmological image to the mean $S_{2}$ scattering coefficients extracted from 1000 realisations of the same input image where the phase information has been randomised. This randomisation step destroys the non-Gaussian information in the original image in which case measuring an excess above unity in the ratio of these scattering coefficients cleanly isolates the non-Gaussian signal.

Next, we demonstrated the application of this approach for realistic mock images of the 21-cm signal with the SKA. In particular, we focussed on the planned SKA imaging survey consisting of a 1000~hr observation spanning 100 square degrees on the sky. For this, we considered two approaches to model the astrophysical foregrounds, one assuming all foregrounds can be perfectly removal and the second assuming foreground avoidance, where all foreground wedge modes are removed from the image. These two approaches act as the two extremes of what should be capable with the SKA. Because the signal can be faint, we also considered two metrics to quantify the detected non-Gaussianity. First, conservatively we measure the mean signal across all 21-cm images covering 100 square degrees. Second, we measure the maximum signal from any one image, in effect focussing on outliers in our sample which have larger than average non-Gaussianity.

We explored the detectability of the non-Gaussian signal at only two observational frequencies, 150~MHz ($z\sim8.5$) and 177~MHz ($z\sim7$). This choice allowed us to focus on understanding and interpreting the method itself rather than complicating matters by attempting to distinguish between astrophysical models. For model and/or parameter inference, the signal needs to be recovered over a broad frequency range, which was studied in detail in \citet{Greig:WST}. For our fiducial model we found that:
\begin{itemize}
\item for foreground removal we could detect non-Gaussianity at S/N~$\gtrsim8$~(5) from the maximum (mean) signal at 177~MHz which corresponds to the latter stages of reionisation, where the signal is expected to be strongest (\avenf~$\sim0.25$). This reduced to S/N~$\gtrsim5$~(3) at 150~MHz just prior to the midpoint (\avenf~$\sim0.6$).
\item under foreground avoidance we recovered notably lower significance detections with S/N~$\gtrsim3$~(1) at 177~MHz and S/N~$\gtrsim2$~(0.5) at 150~MHz. 
\end{itemize}
Note however that for our fiducial model the non-Gaussianity will be larger during the more advanced stages of reionisation (\avenf~$\lesssim0.25$), thus considerably stronger detections would be expected at lower frequencies. Thus, even under foreground avoidance the SKA will yield statistically significant detections of 21-cm non-Gaussianity using the WST.

Finally, at the same two observational frequencies, assuming only foreground avoidance, we provide a quick demonstration of the detectability of the non-Gaussian signal across three other astrophysical models: cold reionisation, extended reionisation and reionisation driven by larger, more biased galaxies. Both cold reionsation and reionisation driven by large halos produce larger amplitude 21-cm brightness temperature contrasts resulting in stronger detections of the non-Gaussianity than our fiducial model: S/N~$\gtrsim13$~(5) using the maximum signal at 177 (150)~MHz for cold reionisation and S/N~$\gtrsim8$~(3) for large haloes. For our extended reionisation model we detect a fainter signal of S/N~$\gtrsim2$~(1.5). Note however that comparing the detectability at a fixed frequency is not a fair comparison as each model may be in a different stage of reionisation and thus may not be where the non-Gaussianity is expected to be maximum. This reinforces the need to consider multiple redshifts to differentiate between astrophysical models.

Importantly, our treatment for the astrophysical foregrounds assumes a pristine cosmological signal, that is, no residual systematics or artefacts are present within the 21-cm images. However, this will not be the case in practice. These artefacts will appear as non-Gaussian information potentially contaminating the cosmological signal. This however, may be somewhat mitigated by measuring the redshift evolution of the non-Gaussian signal. Residual artefacts and systematics should not evolve with redshift and thus may be differentiable from the cosmic signal over time. Further, they may only contaminate certain $S_{2}$ coefficients, leaving others relatively unaffected. We shall explore this in more detail in future work.

\section*{Acknowledgements}

We thank Cathryn Trott and Sihao Cheng for insightful comment and discussions related to this work. Parts of this research were supported by the Australian Research Council Centre of Excellence for All Sky Astrophysics in 3 Dimensions (ASTRO 3D), through project number CE170100013. Y.S.T. acknowledges financial support from the Australian Research Council through DECRA Fellowship DE220101520. Parts of this work were performed on the OzSTAR national facility at Swinburne University of Technology. OzSTAR is funded by Swinburne University of Technology.

\section*{Data Availability}

The data underlying this article will be shared on reasonable request to the corresponding author.

\bibliography{Papers}

\begin{thebibliography}{}
\makeatletter
\relax
\def\mn@urlcharsother{\let\do\@makeother \do\$\do\&\do\#\do\^\do\_\do\%\do\~}
\def\mn@doi{\begingroup\mn@urlcharsother \@ifnextchar [ {\mn@doi@}
  {\mn@doi@[]}}
\def\mn@doi@[#1]#2{\def\@tempa{#1}\ifx\@tempa\@empty \href
  {http://dx.doi.org/#2} {doi:#2}\else \href {http://dx.doi.org/#2} {#1}\fi
  \endgroup}
\def\mn@eprint#1#2{\mn@eprint@#1:#2::\@nil}
\def\mn@eprint@arXiv#1{\href {http://arxiv.org/abs/#1} {{\tt arXiv:#1}}}
\def\mn@eprint@dblp#1{\href {http://dblp.uni-trier.de/rec/bibtex/#1.xml}
  {dblp:#1}}
\def\mn@eprint@#1:#2:#3:#4\@nil{\def\@tempa {#1}\def\@tempb {#2}\def\@tempc
  {#3}\ifx \@tempc \@empty \let \@tempc \@tempb \let \@tempb \@tempa \fi \ifx
  \@tempb \@empty \def\@tempb {arXiv}\fi \@ifundefined
  {mn@eprint@\@tempb}{\@tempb:\@tempc}{\expandafter \expandafter \csname
  mn@eprint@\@tempb\endcsname \expandafter{\@tempc}}}

\bibitem[\protect\citeauthoryear{{Allys}, {Levrier}, {Zhang}, {Colling},
  {Regaldo-Saint Blancard}, {Boulanger}, {Hennebelle}  \& {Mallat}}{{Allys}
  et~al.}{2019}]{Allys:2019}
{Allys} E.,  {Levrier} F.,  {Zhang} S.,  {Colling} C.,  {Regaldo-Saint
  Blancard} B.,  {Boulanger} F.,  {Hennebelle} P.,   {Mallat} S.,  2019,
  \mn@doi [\aap] {10.1051/0004-6361/201834975}, \href
  {https://ui.adsabs.harvard.edu/abs/2019A&A...629A.115A} {629, A115}

\bibitem[\protect\citeauthoryear{{Allys}, {Marchand}, {Cardoso},
  {Villaescusa-Navarro}, {Ho}  \& {Mallat}}{{Allys} et~al.}{2020}]{Allys:2020}
{Allys} E.,  {Marchand} T.,  {Cardoso} J.~F.,  {Villaescusa-Navarro} F.,  {Ho}
  S.,   {Mallat} S.,  2020, \mn@doi [\prd] {10.1103/PhysRevD.102.103506}, \href
  {https://ui.adsabs.harvard.edu/abs/2020PhRvD.102j3506A} {102, 103506}

\bibitem[\protect\citeauthoryear{Alvarez \& Abel}{Alvarez \&
  Abel}{2012}]{Alvarez:2012p1930}
Alvarez M.~A.,  Abel T.,  2012, ApJ, 747, 126

\bibitem[\protect\citeauthoryear{{Bag}, {Mondal}, {Sarkar}, {Bharadwaj},
  {Choudhury}  \& {Sahni}}{{Bag} et~al.}{2019}]{Bag:2019}
{Bag} S.,  {Mondal} R.,  {Sarkar} P.,  {Bharadwaj} S.,  {Choudhury} T.~R.,
  {Sahni} V.,  2019, \mn@doi [\mnras] {10.1093/mnras/stz532}, \href
  {https://ui.adsabs.harvard.edu/abs/2019MNRAS.485.2235B} {485, 2235}

\bibitem[\protect\citeauthoryear{{Banet}, {Barkana}, {Fialkov}  \&
  {Guttman}}{{Banet} et~al.}{2021}]{Banet:2021}
{Banet} A.,  {Barkana} R.,  {Fialkov} A.,   {Guttman} O.,  2021, \mn@doi
  [\mnras] {10.1093/mnras/stab318}, \href
  {https://ui.adsabs.harvard.edu/abs/2021MNRAS.503.1221B} {503, 1221}

\bibitem[\protect\citeauthoryear{{Bianco}, {Giri}, {Iliev}  \&
  {Mellema}}{{Bianco} et~al.}{2021}]{Bianco:2021}
{Bianco} M.,  {Giri} S.~K.,  {Iliev} I.~T.,   {Mellema} G.,  2021, \mn@doi
  [\mnras] {10.1093/mnras/stab1518}, \href
  {https://ui.adsabs.harvard.edu/abs/2021MNRAS.505.3982B} {505, 3982}

\bibitem[\protect\citeauthoryear{{Chege}, {Jordan}, {Lynch}, {Line}  \&
  {Trott}}{{Chege} et~al.}{2021}]{Chege:2021}
{Chege} J.~K.,  {Jordan} C.~H.,  {Lynch} C.,  {Line} J.~L.~B.,   {Trott} C.~M.,
   2021, \mn@doi [\pasa] {10.1017/pasa.2021.22}, \href
  {https://ui.adsabs.harvard.edu/abs/2021PASA...38...28C} {38, e028}

\bibitem[\protect\citeauthoryear{{Chen}, {Xu}, {Wang}  \& {Chen}}{{Chen}
  et~al.}{2019}]{Chen:2019}
{Chen} Z.,  {Xu} Y.,  {Wang} Y.,   {Chen} X.,  2019, \mn@doi [\apj]
  {10.3847/1538-4357/ab43e6}, \href
  {https://ui.adsabs.harvard.edu/abs/2019ApJ...885...23C} {885, 23}

\bibitem[\protect\citeauthoryear{{Cheng} \& {M{\'e}nard}}{{Cheng} \&
  {M{\'e}nard}}{2021}]{Cheng:2021}
{Cheng} S.,  {M{\'e}nard} B.,  2021, \mn@doi [\mnras] {10.1093/mnras/stab2102},
  \href {https://ui.adsabs.harvard.edu/abs/2021MNRAS.507.1012C} {507, 1012}

\bibitem[\protect\citeauthoryear{{Cheng}, {Ting}, {M{\'e}nard}  \&
  {Bruna}}{{Cheng} et~al.}{2020}]{Cheng:2020}
{Cheng} S.,  {Ting} Y.-S.,  {M{\'e}nard} B.,   {Bruna} J.,  2020, \mn@doi
  [\mnras] {10.1093/mnras/staa3165}, \href
  {https://ui.adsabs.harvard.edu/abs/2020MNRAS.499.5902C} {499, 5902}

\bibitem[\protect\citeauthoryear{Datta, Bowman  \& Carilli}{Datta
  et~al.}{2010}]{Datta:2010p2792}
Datta A.,  Bowman J.~D.,   Carilli C.~L.,  2010, ApJ, 724, 526

\bibitem[\protect\citeauthoryear{DeBoer et~al.}{DeBoer
  et~al.}{2017}]{DeBoer:2017p6740}
DeBoer D.~R.,  et~al., 2017, \mn@doi [PASP] {10.1088/1538-3873/129/974/045001},
  \href {https://ui.adsabs.harvard.edu/abs/2017PASP..129d5001D} {129, 045001}

\bibitem[\protect\citeauthoryear{{Doussot}, {Eames}  \& {Semelin}}{{Doussot}
  et~al.}{2019}]{Doussot:2019}
{Doussot} A.,  {Eames} E.,   {Semelin} B.,  2019, \mn@doi [\mnras]
  {10.1093/mnras/stz2429}, \href
  {https://ui.adsabs.harvard.edu/abs/2019MNRAS.490..371D} {490, 371}

\bibitem[\protect\citeauthoryear{{Eastwood} et~al.}{{Eastwood}
  et~al.}{2019}]{Eastwood:2019}
{Eastwood} M.~W.,  et~al., 2019, \mn@doi [AJ] {10.3847/1538-3881/ab2629}, \href
  {https://ui.adsabs.harvard.edu/abs/2019AJ....158...84E} {158, 84}

\bibitem[\protect\citeauthoryear{{Elbers} \& {van de Weygaert}}{{Elbers} \&
  {van de Weygaert}}{2019}]{Elbers:2019}
{Elbers} W.,  {van de Weygaert} R.,  2019, \mn@doi [\mnras]
  {10.1093/mnras/stz908}, \href
  {https://ui.adsabs.harvard.edu/abs/2019MNRAS.486.1523E} {486, 1523}

\bibitem[\protect\citeauthoryear{Field}{Field}{1958}]{Field:1958p1}
Field G.~B.,  1958, \mn@doi [Proc. Inst. Radio Eng.]
  {10.1109/JRPROC.1958.286741}, \href
  {https://ui.adsabs.harvard.edu/abs/1958PIRE...46..240F} {46, 240}

\bibitem[\protect\citeauthoryear{Fragos et~al.}{Fragos
  et~al.}{2013}]{Fragos:2013p6529}
Fragos T.,  et~al., 2013, \mn@doi [ApJ] {10.1088/0004-637X/764/1/41}, \href
  {https://ui.adsabs.harvard.edu/abs/2013ApJ...764...41F} {764, 41}

\bibitem[\protect\citeauthoryear{Furlanetto, Zaldarriaga  \&
  Hernquist}{Furlanetto et~al.}{2004}]{Furlanetto:2004p123}
Furlanetto S.~R.,  Zaldarriaga M.,   Hernquist L.,  2004, \mn@doi [ApJ]
  {10.1086/423025}, \href
  {https://ui.adsabs.harvard.edu/abs/2004ApJ...613....1F} {613, 1}

\bibitem[\protect\citeauthoryear{Furlanetto, Oh  \& Briggs}{Furlanetto
  et~al.}{2006}]{Furlanetto:2006p209}
Furlanetto S.~R.,  Oh S.~P.,   Briggs F.~H.,  2006, \mn@doi [Phys. Rep.]
  {10.1016/j.physrep.2006.08.002}, \href
  {https://ui.adsabs.harvard.edu/abs/2006PhR...433..181F} {433, 181}

\bibitem[\protect\citeauthoryear{{Gazagnes}, {Koopmans}  \&
  {Wilkinson}}{{Gazagnes} et~al.}{2021}]{Gazagnes:2021}
{Gazagnes} S.,  {Koopmans} L. V.~E.,   {Wilkinson} M. H.~F.,  2021, \mn@doi
  [\mnras] {10.1093/mnras/stab107}, \href
  {https://ui.adsabs.harvard.edu/abs/2021MNRAS.502.1816G} {502, 1816}

\bibitem[\protect\citeauthoryear{{Gillet}, {Mesinger}, {Greig}, {Liu}  \&
  {Ucci}}{{Gillet} et~al.}{2019}]{Gillet:2019}
{Gillet} N.,  {Mesinger} A.,  {Greig} B.,  {Liu} A.,   {Ucci} G.,  2019,
  \mn@doi [\mnras] {10.1093/mnras/stz010}, \href
  {https://ui.adsabs.harvard.edu/abs/2019MNRAS.484..282G} {484, 282}

\bibitem[\protect\citeauthoryear{{Giri} \& {Mellema}}{{Giri} \&
  {Mellema}}{2021}]{Giri:2021}
{Giri} S.~K.,  {Mellema} G.,  2021, \mn@doi [\mnras] {10.1093/mnras/stab1320},
  \href {https://ui.adsabs.harvard.edu/abs/2021MNRAS.505.1863G} {505, 1863}

\bibitem[\protect\citeauthoryear{{Giri}, {Mellema}, {Dixon}  \& {Iliev}}{{Giri}
  et~al.}{2018a}]{Giri:2018a}
{Giri} S.~K.,  {Mellema} G.,  {Dixon} K.~L.,   {Iliev} I.~T.,  2018a, \mn@doi
  [\mnras] {10.1093/mnras/stx2539}, \href
  {https://ui.adsabs.harvard.edu/abs/2018MNRAS.473.2949G} {473, 2949}

\bibitem[\protect\citeauthoryear{{Giri}, {Mellema}  \& {Ghara}}{{Giri}
  et~al.}{2018b}]{Giri:2018b}
{Giri} S.~K.,  {Mellema} G.,   {Ghara} R.,  2018b, \mn@doi [\mnras]
  {10.1093/mnras/sty1786}, \href
  {https://ui.adsabs.harvard.edu/abs/2018MNRAS.479.5596G} {479, 5596}

\bibitem[\protect\citeauthoryear{{Giri}, {D'Aloisio}, {Mellema}, {Komatsu},
  {Ghara}  \& {Majumdar}}{{Giri} et~al.}{2019}]{Giri:2019a}
{Giri} S.~K.,  {D'Aloisio} A.,  {Mellema} G.,  {Komatsu} E.,  {Ghara} R.,
  {Majumdar} S.,  2019, \mn@doi [\jcap] {10.1088/1475-7516/2019/02/058}, \href
  {https://ui.adsabs.harvard.edu/abs/2019JCAP...02..058G} {2019, 058}

\bibitem[\protect\citeauthoryear{Gnedin \& Ostriker}{Gnedin \&
  Ostriker}{1997}]{Gnedin:1997p4494}
Gnedin N.~Y.,  Ostriker J.~P.,  1997, ApJ, 486, 581

\bibitem[\protect\citeauthoryear{Gnedin \& Shaver}{Gnedin \&
  Shaver}{2004}]{Gnedin:2004p4481}
Gnedin N.~Y.,  Shaver P.~A.,  2004, ApJ, 608, 611

\bibitem[\protect\citeauthoryear{{Gorce}, {Hutter}  \& {Pritchard}}{{Gorce}
  et~al.}{2021}]{Gorce:2021}
{Gorce} A.,  {Hutter} A.,   {Pritchard} J.~R.,  2021, \mn@doi [\aap]
  {10.1051/0004-6361/202140515}, \href
  {https://ui.adsabs.harvard.edu/abs/2021A&A...653A..58G} {653, A58}

\bibitem[\protect\citeauthoryear{Greig \& Mesinger}{Greig \&
  Mesinger}{2015}]{Greig:2015p3675}
Greig B.,  Mesinger A.,  2015, \mn@doi [MNRAS] {10.1093/mnras/stv571}, \href
  {https://ui.adsabs.harvard.edu/abs/2015MNRAS.449.4246G} {449, 4246}

\bibitem[\protect\citeauthoryear{{Greig}, {Ting}  \& {Kaurov}}{{Greig}
  et~al.}{2022}]{Greig:WST}
{Greig} B.,  {Ting} Y.-S.,   {Kaurov} A.~A.,  2022, \mn@doi [\mnras]
  {10.1093/mnras/stac977}, \href
  {https://ui.adsabs.harvard.edu/abs/2022MNRAS.513.1719G} {513, 1719}

\bibitem[\protect\citeauthoryear{Gupta et~al.}{Gupta et~al.}{2017}]{Gupta:2017}
Gupta Y.,  et~al., 2017, Current Science, \href
  {https://ui.adsabs.harvard.edu/abs/2017CSci..113..707G} {113, 707}

\bibitem[\protect\citeauthoryear{{Hassan}, {Liu}, {Kohn}  \& {La
  Plante}}{{Hassan} et~al.}{2019}]{Hassan:2019}
{Hassan} S.,  {Liu} A.,  {Kohn} S.,   {La Plante} P.,  2019, \mn@doi [\mnras]
  {10.1093/mnras/sty3282}, \href
  {https://ui.adsabs.harvard.edu/abs/2019MNRAS.483.2524H} {483, 2524}

\bibitem[\protect\citeauthoryear{{Hassan}, {Andrianomena}  \&
  {Doughty}}{{Hassan} et~al.}{2020}]{Hassan:2020}
{Hassan} S.,  {Andrianomena} S.,   {Doughty} C.,  2020, \mn@doi [\mnras]
  {10.1093/mnras/staa1151}, \href
  {https://ui.adsabs.harvard.edu/abs/2020MNRAS.494.5761H} {494, 5761}

\bibitem[\protect\citeauthoryear{Hutter, Dayal, Yepes, Gottl{\"o}ber, Legrand
  \& Ucci}{Hutter et~al.}{2020}]{Hutter:2020}
Hutter A.,  Dayal P.,  Yepes G.,  Gottl{\"o}ber S.,  Legrand L.,   Ucci G.,
  2020, arXiv e-prints, \href
  {https://ui.adsabs.harvard.edu/abs/2020arXiv200408401H} {p. arXiv:2004.08401}

\bibitem[\protect\citeauthoryear{{Jordan} et~al.,}{{Jordan}
  et~al.}{2017}]{Jordan:2017}
{Jordan} C.~H.,  et~al., 2017, \mn@doi [\mnras] {10.1093/mnras/stx1797}, \href
  {https://ui.adsabs.harvard.edu/abs/2017MNRAS.471.3974J} {471, 3974}

\bibitem[\protect\citeauthoryear{{Kakiichi} et~al.,}{{Kakiichi}
  et~al.}{2017}]{Kakiichi:2017}
{Kakiichi} K.,  et~al., 2017, \mn@doi [\mnras] {10.1093/mnras/stx1568}, \href
  {https://ui.adsabs.harvard.edu/abs/2017MNRAS.471.1936K} {471, 1936}

\bibitem[\protect\citeauthoryear{{Kamran}, {Ghara}, {Majumdar}, {Mondal},
  {Mellema}, {Bharadwaj}, {Pritchard}  \& {Iliev}}{{Kamran}
  et~al.}{2021}]{Kamran:2021}
{Kamran} M.,  {Ghara} R.,  {Majumdar} S.,  {Mondal} R.,  {Mellema} G.,
  {Bharadwaj} S.,  {Pritchard} J.~R.,   {Iliev} I.~T.,  2021, \mn@doi [\mnras]
  {10.1093/mnras/stab216}, \href
  {https://ui.adsabs.harvard.edu/abs/2021MNRAS.502.3800K} {502, 3800}

\bibitem[\protect\citeauthoryear{{Kapahtia}, {Chingangbam}  \&
  {Appleby}}{{Kapahtia} et~al.}{2019}]{Kapahtia:2019}
{Kapahtia} A.,  {Chingangbam} P.,   {Appleby} S.,  2019, \mn@doi [\jcap]
  {10.1088/1475-7516/2019/09/053}, \href
  {https://ui.adsabs.harvard.edu/abs/2019JCAP...09..053K} {2019, 053}

\bibitem[\protect\citeauthoryear{{Kapahtia}, {Chingangbam}, {Ghara}, {Appleby}
  \& {Choudhury}}{{Kapahtia} et~al.}{2021}]{Kapahtia:2021}
{Kapahtia} A.,  {Chingangbam} P.,  {Ghara} R.,  {Appleby} S.,   {Choudhury}
  T.~R.,  2021, \mn@doi [\jcap] {10.1088/1475-7516/2021/05/026}, \href
  {https://ui.adsabs.harvard.edu/abs/2021JCAP...05..026K} {2021, 026}

\bibitem[\protect\citeauthoryear{{Koopmans} et~al.}{{Koopmans}
  et~al.}{2015}]{Koopmans:2015}
{Koopmans} L.,  et~al., 2015, in Advancing Astrophysics with the Square
  Kilometre Array (AASKA14).  (\mn@eprint {arXiv} {1505.07568})

\bibitem[\protect\citeauthoryear{{Kubota}, {Yoshiura}, {Shimabukuro}  \&
  {Takahashi}}{{Kubota} et~al.}{2016}]{Kubota:2016}
{Kubota} K.,  {Yoshiura} S.,  {Shimabukuro} H.,   {Takahashi} K.,  2016,
  \mn@doi [\pasj] {10.1093/pasj/psw059}, \href
  {https://ui.adsabs.harvard.edu/abs/2016PASJ...68...61K} {68, 61}

\bibitem[\protect\citeauthoryear{{Kwon}, {Hong}  \& {Park}}{{Kwon}
  et~al.}{2020}]{Kwon:2020}
{Kwon} Y.,  {Hong} S.~E.,   {Park} I.,  2020, \mn@doi [Journal of Korean
  Physical Society] {10.3938/jkps.77.49}, \href
  {https://ui.adsabs.harvard.edu/abs/2020JKPS...77...49K} {77, 49}

\bibitem[\protect\citeauthoryear{{La Plante} \& {Ntampaka}}{{La Plante} \&
  {Ntampaka}}{2019}]{LaPlante:2019}
{La Plante} P.,  {Ntampaka} M.,  2019, \mn@doi [\apj]
  {10.3847/1538-4357/ab2983}, \href
  {https://ui.adsabs.harvard.edu/abs/2019ApJ...880..110L} {880, 110}

\bibitem[\protect\citeauthoryear{Liu, Parsons  \& Trott}{Liu
  et~al.}{2014a}]{Liu:2014p3465}
Liu A.,  Parsons A.~R.,   Trott C.~M.,  2014a, Phys. Rev. D, 90, 023018

\bibitem[\protect\citeauthoryear{Liu, Parsons  \& Trott}{Liu
  et~al.}{2014b}]{Liu:2014p3466}
Liu A.,  Parsons A.~R.,   Trott C.~M.,  2014b, Phys. Rev. D, 90, 023019

\bibitem[\protect\citeauthoryear{Madau, Meiksin  \& Rees}{Madau
  et~al.}{1997}]{Madau:1997p4479}
Madau P.,  Meiksin A.,   Rees M.~J.,  1997, ApJ, 475, 429

\bibitem[\protect\citeauthoryear{{Majumdar}, {Pritchard}, {Mondal},
  {Watkinson}, {Bharadwaj}  \& {Mellema}}{{Majumdar}
  et~al.}{2018}]{Majumdar:2018}
{Majumdar} S.,  {Pritchard} J.~R.,  {Mondal} R.,  {Watkinson} C.~A.,
  {Bharadwaj} S.,   {Mellema} G.,  2018, \mn@doi [\mnras]
  {10.1093/mnras/sty535}, \href
  {https://ui.adsabs.harvard.edu/abs/2018MNRAS.476.4007M} {476, 4007}

\bibitem[\protect\citeauthoryear{{Majumdar}, {Kamran}, {Pritchard}, {Mondal},
  {Mazumdar}, {Bharadwaj}  \& {Mellema}}{{Majumdar}
  et~al.}{2020}]{Majumdar:2020}
{Majumdar} S.,  {Kamran} M.,  {Pritchard} J.~R.,  {Mondal} R.,  {Mazumdar} A.,
  {Bharadwaj} S.,   {Mellema} G.,  2020, \mn@doi [\mnras]
  {10.1093/mnras/staa3168}, \href
  {https://ui.adsabs.harvard.edu/abs/2020MNRAS.499.5090M} {499, 5090}

\bibitem[\protect\citeauthoryear{{Mallat}}{{Mallat}}{2012}]{Mallat:2012}
{Mallat} S.,  2012, \mn@doi [Communications on Pure and Applied Mathematics]
  {10.1002/cpa.21413}, \href {https://arxiv.org/abs/1101.2286} {65, 1331}

\bibitem[\protect\citeauthoryear{{Mangena}, {Hassan}  \& {Santos}}{{Mangena}
  et~al.}{2020}]{Mangena:2020}
{Mangena} T.,  {Hassan} S.,   {Santos} M.~G.,  2020, \mn@doi [\mnras]
  {10.1093/mnras/staa750}, \href
  {https://ui.adsabs.harvard.edu/abs/2020MNRAS.494..600M} {494, 600}

\bibitem[\protect\citeauthoryear{Mellema et~al.}{Mellema
  et~al.}{2013}]{Mellema:2013p2975}
Mellema G.,  et~al., 2013, \mn@doi [Exp. Astron.] {10.1007/s10686-013-9334-5},
  \href {https://ui.adsabs.harvard.edu/abs/2013ExA....36..235M} {36, 235}

\bibitem[\protect\citeauthoryear{Mesinger \& Furlanetto}{Mesinger \&
  Furlanetto}{2007}]{Mesinger:2007p122}
Mesinger A.,  Furlanetto S.,  2007, \mn@doi [ApJ] {10.1086/521806}, \href
  {https://ui.adsabs.harvard.edu/abs/2007ApJ...669..663M} {669, 663}

\bibitem[\protect\citeauthoryear{Mesinger, Furlanetto  \& Cen}{Mesinger
  et~al.}{2011}]{Mesinger:2011p1123}
Mesinger A.,  Furlanetto S.,   Cen R.,  2011, \mn@doi [MNRAS]
  {10.1111/j.1365-2966.2010.17731.x}, \href
  {https://ui.adsabs.harvard.edu/abs/2011MNRAS.411..955M} {411, 955}

\bibitem[\protect\citeauthoryear{Mesinger, McQuinn  \& Spergel}{Mesinger
  et~al.}{2012}]{Mesinger:2012p1131}
Mesinger A.,  McQuinn M.,   Spergel D.~N.,  2012, MNRAS, 422, 1403

\bibitem[\protect\citeauthoryear{Mesinger, Ewall-Wice  \& Hewitt}{Mesinger
  et~al.}{2014}]{Mesinger:2014p244}
Mesinger A.,  Ewall-Wice A.,   Hewitt J.,  2014, \mn@doi [MNRAS]
  {10.1093/mnras/stu125}, \href
  {https://ui.adsabs.harvard.edu/abs/2014MNRAS.439.3262M} {439, 3262}

\bibitem[\protect\citeauthoryear{Mineo, Gilfanov  \& Sunyaev}{Mineo
  et~al.}{2012}]{Mineo:2012p6282}
Mineo S.,  Gilfanov M.,   Sunyaev R.,  2012, \mn@doi [MNRAS]
  {10.1111/j.1365-2966.2011.19862.x}, \href
  {https://ui.adsabs.harvard.edu/abs/2012MNRAS.419.2095M} {419, 2095}

\bibitem[\protect\citeauthoryear{Morales \& Wyithe}{Morales \&
  Wyithe}{2010}]{Morales:2010p1274}
Morales M.~F.,  Wyithe J. S.~B.,  2010, \mn@doi [ARA\&A]
  {10.1146/annurev-astro-081309-130936}, \href
  {https://ui.adsabs.harvard.edu/abs/2010ARA&A..48..127M} {48, 127}

\bibitem[\protect\citeauthoryear{Morales, Hazelton, Sullivan  \&
  Beardsley}{Morales et~al.}{2012}]{Morales:2012p2828}
Morales M.~F.,  Hazelton B.,  Sullivan I.,   Beardsley A.,  2012, ApJ, 752, 137

\bibitem[\protect\citeauthoryear{{Mu{\~n}oz}, {Qin}, {Mesinger}, {Murray},
  {Greig}  \& {Mason}}{{Mu{\~n}oz} et~al.}{2022}]{Munoz:2022}
{Mu{\~n}oz} J.~B.,  {Qin} Y.,  {Mesinger} A.,  {Murray} S.~G.,  {Greig} B.,
  {Mason} C.,  2022, \mn@doi [\mnras] {10.1093/mnras/stac185}, \href
  {https://ui.adsabs.harvard.edu/abs/2022MNRAS.511.3657M} {511, 3657}

\bibitem[\protect\citeauthoryear{{Murray} \& {Trott}}{{Murray} \&
  {Trott}}{2018}]{Murray:2018}
{Murray} S.~G.,  {Trott} C.~M.,  2018, \mn@doi [ApJ]
  {10.3847/1538-4357/aaebfa}, \href
  {https://ui.adsabs.harvard.edu/abs/2018ApJ...869...25M} {869, 25}

\bibitem[\protect\citeauthoryear{{Murray}, {Greig}, {Mesinger}, {Mu{\~n}oz},
  {Qin}, {Park}  \& {Watkinson}}{{Murray} et~al.}{2020}]{Murray:2020}
{Murray} S.,  {Greig} B.,  {Mesinger} A.,  {Mu{\~n}oz} J.,  {Qin} Y.,  {Park}
  J.,   {Watkinson} C.,  2020, \mn@doi [The Journal of Open Source Software]
  {10.21105/joss.02582}, \href
  {https://ui.adsabs.harvard.edu/abs/2020JOSS....5.2582M} {5, 2582}

\bibitem[\protect\citeauthoryear{Pacucci, Mesinger, Mineo  \& Ferrara}{Pacucci
  et~al.}{2014}]{Pacucci:2014p4323}
Pacucci F.,  Mesinger A.,  Mineo S.,   Ferrara A.,  2014, \mn@doi [MNRAS]
  {10.1093/mnras/stu1240}, \href
  {https://ui.adsabs.harvard.edu/abs/2014MNRAS.443..678P} {443, 678}

\bibitem[\protect\citeauthoryear{{Park}, {Mesinger}, {Greig}  \&
  {Gillet}}{{Park} et~al.}{2019}]{Park:2019}
{Park} J.,  {Mesinger} A.,  {Greig} B.,   {Gillet} N.,  2019, \mn@doi [MNRAS]
  {10.1093/mnras/stz032}, \href
  {https://ui.adsabs.harvard.edu/abs/2019MNRAS.484..933P} {484, 933}

\bibitem[\protect\citeauthoryear{Parsons et~al.}{Parsons
  et~al.}{2010}]{Parsons:2010p3000}
Parsons A.~R.,  et~al., 2010, \mn@doi [AJ] {10.1088/0004-6256/139/4/1468},
  \href {https://ui.adsabs.harvard.edu/abs/2010AJ....139.1468P} {139, 1468}

\bibitem[\protect\citeauthoryear{Parsons, Pober, Aguirre, Carilli, Jacobs  \&
  Moore}{Parsons et~al.}{2012}]{Parsons:2012p2833}
Parsons A.~R.,  Pober J.~C.,  Aguirre J.~E.,  Carilli C.~L.,  Jacobs D.~C.,
  Moore D.~F.,  2012, ApJ, 756, 165

\bibitem[\protect\citeauthoryear{Parsons et~al.}{Parsons
  et~al.}{2014}]{Parsons:2014p781}
Parsons A.~R.,  et~al., 2014, \mn@doi [ApJ] {10.1088/0004-637X/788/2/106},
  \href {https://ui.adsabs.harvard.edu/abs/2014ApJ...788..106P} {788, 106}

\bibitem[\protect\citeauthoryear{{Planck Collaboration} et~al.}{{Planck
  Collaboration} et~al.}{2020}]{Planck:2020}
{Planck Collaboration} et~al., 2020, \mn@doi [\aap]
  {10.1051/0004-6361/201833910}, \href
  {https://ui.adsabs.harvard.edu/abs/2020A&A...641A...6P} {641, A6}

\bibitem[\protect\citeauthoryear{Pober et~al.}{Pober
  et~al.}{2013}]{Pober:2013p41}
Pober J.~C.,  et~al., 2013, AJ, 145, 65

\bibitem[\protect\citeauthoryear{Pober et~al.}{Pober
  et~al.}{2014}]{Pober:2014p35}
Pober J.~C.,  et~al., 2014, ApJ, 782, 66

\bibitem[\protect\citeauthoryear{Pober et~al.}{Pober
  et~al.}{2016}]{Pober:2016p7301}
Pober J.~C.,  et~al., 2016, ApJ, 819, 8

\bibitem[\protect\citeauthoryear{{Prelogovi{\'c}}, {Mesinger}, {Murray},
  {Fiameni}  \& {Gillet}}{{Prelogovi{\'c}} et~al.}{2021}]{Prelogovic:2021}
{Prelogovi{\'c}} D.,  {Mesinger} A.,  {Murray} S.,  {Fiameni} G.,   {Gillet}
  N.,  2021, arXiv e-prints, \href
  {https://ui.adsabs.harvard.edu/abs/2021arXiv210700018P} {p. arXiv:2107.00018}

\bibitem[\protect\citeauthoryear{Pritchard \& Loeb}{Pritchard \&
  Loeb}{2012}]{Pritchard:2012p2958}
Pritchard J.~R.,  Loeb A.,  2012, \mn@doi [Rep. Prog. Phys.]
  {10.1088/0034-4885/75/8/086901}, \href
  {https://ui.adsabs.harvard.edu/abs/2012RPPh...75h6901P} {75, 086901}

\bibitem[\protect\citeauthoryear{{Qin}, {Mesinger}, {Bosman}  \& {Viel}}{{Qin}
  et~al.}{2021}]{Qin:2021}
{Qin} Y.,  {Mesinger} A.,  {Bosman} S. E.~I.,   {Viel} M.,  2021, \mn@doi
  [\mnras] {10.1093/mnras/stab1833}, \href
  {https://ui.adsabs.harvard.edu/abs/2021MNRAS.506.2390Q} {506, 2390}

\bibitem[\protect\citeauthoryear{{Rahimi} et~al.,}{{Rahimi}
  et~al.}{2021}]{Rahimi:2021}
{Rahimi} M.,  et~al., 2021, \mn@doi [\mnras] {10.1093/mnras/stab2918}, \href
  {https://ui.adsabs.harvard.edu/abs/2021MNRAS.508.5954R} {508, 5954}

\bibitem[\protect\citeauthoryear{{Regaldo-Saint Blancard}, {Levrier}, {Allys},
  {Bellomi}  \& {Boulanger}}{{Regaldo-Saint Blancard}
  et~al.}{2020}]{Blancard:2020}
{Regaldo-Saint Blancard} B.,  {Levrier} F.,  {Allys} E.,  {Bellomi} E.,
  {Boulanger} F.,  2020, \mn@doi [\aap] {10.1051/0004-6361/202038044}, \href
  {https://ui.adsabs.harvard.edu/abs/2020A&A...642A.217R} {642, A217}

\bibitem[\protect\citeauthoryear{{Salpeter}}{{Salpeter}}{1955}]{Salpeter:1955}
{Salpeter} E.~E.,  1955, \mn@doi [ApJ] {10.1086/145971}, \href
  {https://ui.adsabs.harvard.edu/abs/1955ApJ...121..161S} {121, 161}

\bibitem[\protect\citeauthoryear{{Saydjari}, {Portillo}, {Slepian}, {Kahraman},
  {Burkhart}  \& {Finkbeiner}}{{Saydjari} et~al.}{2021}]{Saydjari:2021}
{Saydjari} A.~K.,  {Portillo} S. K.~N.,  {Slepian} Z.,  {Kahraman} S.,
  {Burkhart} B.,   {Finkbeiner} D.~P.,  2021, \mn@doi [\apj]
  {10.3847/1538-4357/abe46d}, \href
  {https://ui.adsabs.harvard.edu/abs/2021ApJ...910..122S} {910, 122}

\bibitem[\protect\citeauthoryear{Shaver, Windhorst, Madau  \& de Bruyn}{Shaver
  et~al.}{1999}]{Shaver:1999p4549}
Shaver P.~A.,  Windhorst R.~A.,  Madau P.,   de Bruyn A.~G.,  1999, A\&A, 345,
  380

\bibitem[\protect\citeauthoryear{Sheth \& Tormen}{Sheth \&
  Tormen}{1999}]{Sheth:1999p2053}
Sheth R.~K.,  Tormen G.,  1999, \mn@doi [MNRAS]
  {10.1046/j.1365-8711.1999.02692.x}, \href
  {https://ui.adsabs.harvard.edu/abs/1999MNRAS.308..119S} {308, 119}

\bibitem[\protect\citeauthoryear{{Shimabukuro}, {Yoshiura}, {Takahashi},
  {Yokoyama}  \& {Ichiki}}{{Shimabukuro} et~al.}{2015}]{Shimabukuro:2015}
{Shimabukuro} H.,  {Yoshiura} S.,  {Takahashi} K.,  {Yokoyama} S.,   {Ichiki}
  K.,  2015, \mn@doi [\mnras] {10.1093/mnras/stv965}, \href
  {https://ui.adsabs.harvard.edu/abs/2015MNRAS.451..467S} {451, 467}

\bibitem[\protect\citeauthoryear{{Shimabukuro}, {Yoshiura}, {Takahashi},
  {Yokoyama}  \& {Ichiki}}{{Shimabukuro} et~al.}{2016}]{Shimabukuro:2016}
{Shimabukuro} H.,  {Yoshiura} S.,  {Takahashi} K.,  {Yokoyama} S.,   {Ichiki}
  K.,  2016, \mn@doi [\mnras] {10.1093/mnras/stw482}, \href
  {https://ui.adsabs.harvard.edu/abs/2016MNRAS.458.3003S} {458, 3003}

\bibitem[\protect\citeauthoryear{Sobacchi \& Mesinger}{Sobacchi \&
  Mesinger}{2014}]{Sobacchi:2014p1157}
Sobacchi E.,  Mesinger A.,  2014, \mn@doi [MNRAS] {10.1093/mnras/stu377}, \href
  {https://ui.adsabs.harvard.edu/abs/2014MNRAS.440.1662S} {440, 1662}

\bibitem[\protect\citeauthoryear{{Thompson}, {Moran}  \& {Swenson}}{{Thompson}
  et~al.}{2007}]{Thompson2007}
{Thompson} A.~R.,  {Moran} J.~M.,   {Swenson} G.~W.,  2007, {in Interferometry
  and Synthesis in Radio Astronomy. Wiley, New York}

\bibitem[\protect\citeauthoryear{Thyagarajan et~al.}{Thyagarajan
  et~al.}{2013}]{Thyagarajan:2013p2851}
Thyagarajan N.,  et~al., 2013, ApJ, 776, 6

\bibitem[\protect\citeauthoryear{Thyagarajan et~al.}{Thyagarajan
  et~al.}{2015a}]{Thyagarajan:2015p7294}
Thyagarajan N.,  et~al., 2015a, ApJ, 804, 14

\bibitem[\protect\citeauthoryear{Thyagarajan et~al.}{Thyagarajan
  et~al.}{2015b}]{Thyagarajan:2015p7298}
Thyagarajan N.,  et~al., 2015b, ApJL, 807, L28

\bibitem[\protect\citeauthoryear{Tingay et~al.}{Tingay
  et~al.}{2013}]{Tingay:2013p2997}
Tingay S.~J.,  et~al., 2013, \mn@doi [PASA] {10.1017/pasa.2012.007}, \href
  {https://ui.adsabs.harvard.edu/abs/2013PASA...30....7T} {30, 7}

\bibitem[\protect\citeauthoryear{Tozzi, Madau, Meiksin  \& Rees}{Tozzi
  et~al.}{2000}]{Tozzi:2000p4510}
Tozzi P.,  Madau P.,  Meiksin A.,   Rees M.~J.,  2000, ApJ, 528, 597

\bibitem[\protect\citeauthoryear{Trott, Wayth  \& Tingay}{Trott
  et~al.}{2012}]{Trott:2012p2834}
Trott C.~M.,  Wayth R.~B.,   Tingay S.~J.,  2012, ApJ, 757, 101

\bibitem[\protect\citeauthoryear{{Valogiannis} \& {Dvorkin}}{{Valogiannis} \&
  {Dvorkin}}{2021}]{Valogiannis:2021}
{Valogiannis} G.,  {Dvorkin} C.,  2021, arXiv e-prints, \href
  {https://ui.adsabs.harvard.edu/abs/2021arXiv210807821V} {p. arXiv:2108.07821}

\bibitem[\protect\citeauthoryear{{Valogiannis} \& {Dvorkin}}{{Valogiannis} \&
  {Dvorkin}}{2022}]{Valogiannis:2022}
{Valogiannis} G.,  {Dvorkin} C.,  2022, arXiv e-prints, \href
  {https://ui.adsabs.harvard.edu/abs/2022arXiv220413717V} {p. arXiv:2204.13717}

\bibitem[\protect\citeauthoryear{Vedantham, Shankar  \& Subrahmanyan}{Vedantham
  et~al.}{2012}]{Vedantham:2012p2801}
Vedantham H.,  Shankar N.~U.,   Subrahmanyan R.,  2012, ApJ, 745, 176

\bibitem[\protect\citeauthoryear{{Watkinson} \& {Pritchard}}{{Watkinson} \&
  {Pritchard}}{2014}]{Watkinson:2014}
{Watkinson} C.~A.,  {Pritchard} J.~R.,  2014, \mn@doi [\mnras]
  {10.1093/mnras/stu1384}, \href
  {https://ui.adsabs.harvard.edu/abs/2014MNRAS.443.3090W} {443, 3090}

\bibitem[\protect\citeauthoryear{{Watkinson}, {Giri}, {Ross}, {Dixon}, {Iliev},
  {Mellema}  \& {Pritchard}}{{Watkinson} et~al.}{2019}]{Watkinson:2019}
{Watkinson} C.~A.,  {Giri} S.~K.,  {Ross} H.~E.,  {Dixon} K.~L.,  {Iliev}
  I.~T.,  {Mellema} G.,   {Pritchard} J.~R.,  2019, \mn@doi [\mnras]
  {10.1093/mnras/sty2740}, \href
  {https://ui.adsabs.harvard.edu/abs/2019MNRAS.482.2653W} {482, 2653}

\bibitem[\protect\citeauthoryear{Wayth et~al.}{Wayth et~al.}{2018}]{Wayth:2018}
Wayth R.,  et~al., 2018, \mn@doi [\pasa] {10.1017/pasa.2018.37}, \href
  {https://ui.adsabs.harvard.edu/abs/2018PASA...35...33W} {35, 33}

\bibitem[\protect\citeauthoryear{Wouthuysen}{Wouthuysen}{1952}]{Wouthuysen:1952p4321}
Wouthuysen S.~A.,  1952, \mn@doi [AJ] {10.1086/106661}, \href
  {https://ui.adsabs.harvard.edu/abs/1952AJ.....57R..31W} {57, 31}

\bibitem[\protect\citeauthoryear{{Yoshiura}, {Shimabukuro}, {Takahashi},
  {Momose}, {Nakanishi}  \& {Imai}}{{Yoshiura} et~al.}{2015}]{Yoshiura:2015}
{Yoshiura} S.,  {Shimabukuro} H.,  {Takahashi} K.,  {Momose} R.,  {Nakanishi}
  H.,   {Imai} H.,  2015, \mn@doi [\mnras] {10.1093/mnras/stv855}, \href
  {https://ui.adsabs.harvard.edu/abs/2015MNRAS.451..266Y} {451, 266}

\bibitem[\protect\citeauthoryear{{Yoshiura}, {Shimabukuro}, {Takahashi}  \&
  {Matsubara}}{{Yoshiura} et~al.}{2017}]{Yoshiura:2017}
{Yoshiura} S.,  {Shimabukuro} H.,  {Takahashi} K.,   {Matsubara} T.,  2017,
  \mn@doi [\mnras] {10.1093/mnras/stw2701}, \href
  {https://ui.adsabs.harvard.edu/abs/2017MNRAS.465..394Y} {465, 394}

\bibitem[\protect\citeauthoryear{{Zarka}, {Girard}, {Tagger}  \&
  {Denis}}{{Zarka} et~al.}{2012}]{Zarka:2012}
{Zarka} P.,  {Girard} J.~N.,  {Tagger} M.,   {Denis} L.,  2012, in {Boissier}
  S.,  {de Laverny} P.,  {Nardetto} N.,  {Samadi} R.,  {Valls-Gabaud} D.,
  {Wozniak} H.,  eds, SF2A-2012: Proceedings of the Annual meeting of the
  French Society of Astronomy and Astrophysics. pp 687--694

\bibitem[\protect\citeauthoryear{{Zhao}, {Mao}, {Cheng}  \& {Wandelt}}{{Zhao}
  et~al.}{2022}]{Zhao:2022}
{Zhao} X.,  {Mao} Y.,  {Cheng} C.,   {Wandelt} B.~D.,  2022, \mn@doi [\apj]
  {10.3847/1538-4357/ac457d}, \href
  {https://ui.adsabs.harvard.edu/abs/2022ApJ...926..151Z} {926, 151}

\bibitem[\protect\citeauthoryear{van Haarlem et~al.}{van Haarlem
  et~al.}{2013}]{vanHaarlem:2013p200}
van Haarlem M.~P.,  et~al., 2013, \mn@doi [A\&A] {10.1051/0004-6361/201220873},
  \href {https://ui.adsabs.harvard.edu/abs/2013A&A...556A...2V} {556, 2}

\makeatother
\end{thebibliography}

\appendix

\section{Impact of Observing Time} \label{sec:obs_time}

\begin{figure*} 
	\begin{center}
		\includegraphics[trim = 1.3cm 0.7cm 0cm 0.5cm, scale = 0.56]{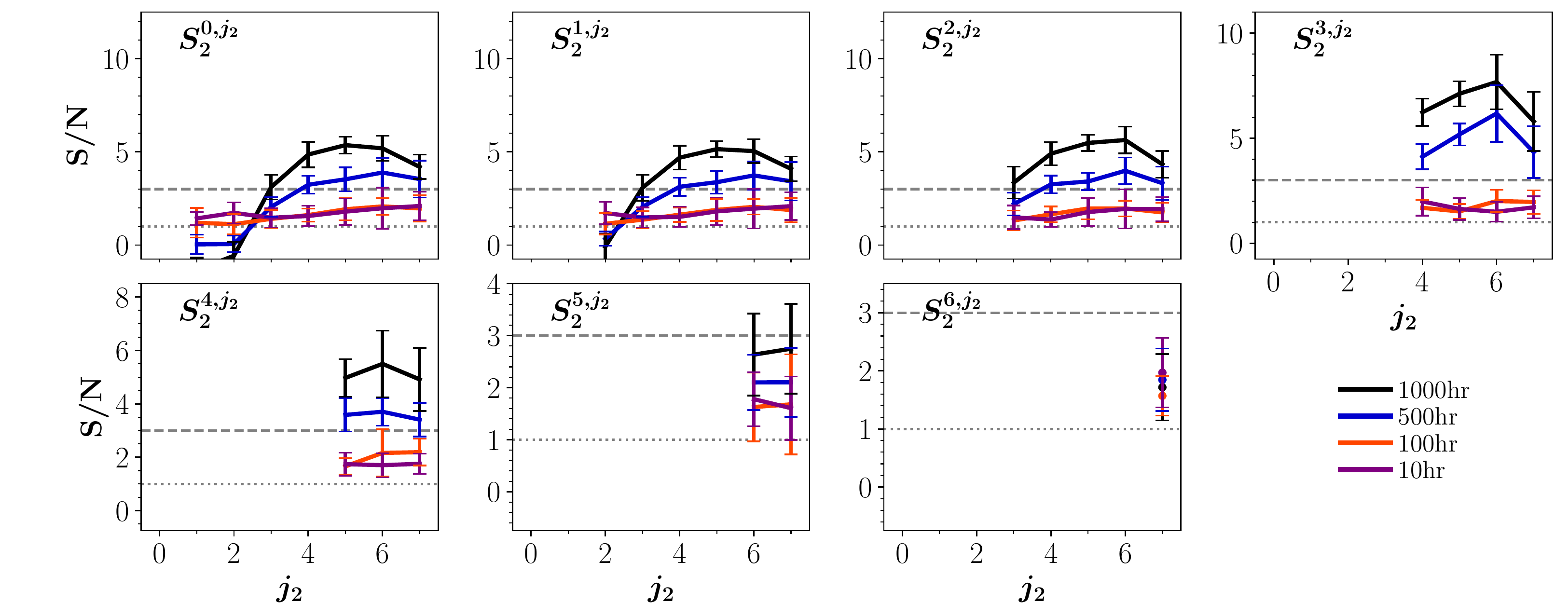}
	\end{center}
\caption[]{The corresponding signal-to-noise (S/N) of the non-Gaussian signal from the second-order scattering coefficients ($S_{2}$) for differing observing times. We consider our fiducial model and perfect foreground removal assuming: 1000~hr (black curves), 500~hr (blue), 100~hr (red) and 10~hr (purple).}
\label{fig:obs_time}
\end{figure*}

All results reported in Section~\ref{sec:results} assume the full 1000~hr observing time with the SKA following the planned deep survey. Here, we explore the impact of different observing times at recovering the non-Gaussianity in the 21-cm signal. For this, we assume our fiducial astrophysical model at 150~MHz adopting perfect foreground removal (to more clearly visualise the impact).

In Figure~\ref{fig:obs_time} we present the S/N of the non-Gaussian signal for the full 1000~hr (black), 500~hr (blue), 100~hr (red) and 10~hr (purple curves), respectively. For both our nominal 1000~hr survey and for 500~hrs, the S/N of the non-Gaussianity remains above three for a large range of $S_{2}$ coefficients, differing by a factor of two between the two observing times. At only 100~hrs, we still recover a S/N of up to $\sim2$, despite the drop by an order of magnitude in observing time. The same is true for only 10~hrs observing time, where despite the considerable drop in observing time (factor of 100) something is still marginally detectable S/N~$\sim1-2$. This implies that this non-Gaussianity is still detectable for shorter observing times, but at notably reduced sensitivities. 

Despite the dropping observing time, the S/N appears to remain at S/N~$\sim1-2$ even for 10~hr and is comparable to those results at 100~hr. This apparent floor, above zero, occurs because we are measuring the maximum signal rather than the mean signal. That is, since we are effectively sampling the outlier of our 10 images, we only require one of our images to have a large enough non-Gaussian signal to detect it. Indeed, this is evident from the amplitude of the error bars between the 10 and 100~hr scenarios. Although the amplitude of the S/N is almost identical between these two, the 68th percentiles on a 10~hr observation are larger than those for the 100~hr observation. In fact, the uncertainties on the 10~hr observing time are the largest of all four scenarios, indicative of increased sensitivity to individual (outlier) images.

\end{document}